\documentclass[%
 reprint,
amsmath,amssymb,
 aps,
]{revtex4-2}

\usepackage[justification=raggedright,singlelinecheck=false]{caption}
\usepackage{graphicx}
\usepackage{dcolumn}
\usepackage{bm}
\usepackage{hyperref}
\usepackage{color}
\usepackage{amsmath,amssymb,amsthm}
\usepackage{chngcntr}
\usepackage{subcaption}
\usepackage{apptools}
\AtAppendix{}

\begin{document}

\preprint{APS/123-QED}

\title{Stochastic resetting in a nonequilibrium environment }

\author{Koushik Goswami}
\email{goswamikoushik10@gmail.com}
\affiliation{Physics Division,
National Center for Theoretical Sciences, 
National Taiwan University, Taipei 106319, Taiwan}

\begin{abstract}
This study examines the dynamics of a tracer particle diffusing in a nonequilibrium medium under stochastic resetting. The nonequilibrium state is induced by harmonic coupling between the tracer and bath particles, generating memory effects with exponential decay in time. We explore the tracer’s behavior under a Poissonian resetting protocol, where resetting does not disturb the bath environment, with a focus on key dynamical behavior  and first-passage properties, both in the presence and absence of an external force. The interplay between coupling strength and diffusivity of bath particle  significantly impacts both the tracer’s relaxation dynamics and search time, with external forces further modulating these effects.
Our analysis identifies distinct hot and cold bath particles based on their diffusivities, revealing that coupling to a hot particle facilitates the searching process, whereas coupling to a cold particle hinders it.  Using a combination of numerical simulations and analytical methods, this study provides a comprehensive framework  for understanding resetting mechanisms in non-Markovian systems, with potential applications to complex environments such as active and viscoelastic media, where memory-driven dynamics and nonequilibrium interactions are significant.
\end{abstract}

\maketitle



\section{Introduction}
The diffusion of a tracer, a classical example of Brownian motion, is a well-established area of research that has served as a cornerstone in advancing our understanding of stochastic processes for over a century \cite{van1992stochastic,frey2005brownian}. In a thermal environment, the erratic motion of a tracer arises from consistent collisions with small, fast-moving bath particles, resulting in instantaneous random forces that lead to  the tracer's fluctuating movement. By assuming the rapid relaxation of bath particles and averaging out these fast degrees of freedom, one can derive the effective dynamics of the tracer, as originally formulated by Paul Langevin and now known as the Langevin equation \cite{Zwanzig2002,Kubo2012}. In this context, the tracer experiences a drag force with a constant drag coefficient, making its future trajectory dependent solely on its current state, rather than its entire history—characteristic of Markovian dynamics \cite{Kubo2012}.

Although Markovian dynamics are relatively simple to study, they fall short in describing the behavior of complex systems, such as transport in complex and viscoelastic media  \cite{Metzler2000,Hofling2013,Ferrer2021}, conformational changes in macromolecules \cite{Kappler2018}, cell migration \cite{Mitterwallner2020}, and protein folding \cite{Makarov2013}, where memory effects—often characterized by generalized friction or a memory kernel—become significant due to the relaxation of all degrees of freedom over comparable timescales. A typical example is the tracer dynamics in a gel, such as the cytoskeletal network composed of cross-linked actin filaments and motor proteins. The activity of these motor proteins generates fluctuations in the network, leading to memory-induced forces that drive the movement of the tracer \cite{Mizuno2007,Godec2014,Joo2022,Theeyancheri2022}.  Similarly, the shape fluctuations of cellular or nuclear membranes are caused by the rearrangements of the underlying network, driven by non-equilibrium fluctuations \cite{Faris2009,Park2010}. These processes are driven out of equilibrium due to the continuous injection of energy into the system, and therefore cannot be fully explained within the framework of equilibrium theory \cite{Crisanti2012,Fodor2014,Hiraiwa2018,Netz2018,Goswami2023}. Such fluctuations are often modeled by coupling the tracer to the bath's degrees of freedom through an interaction potential. An analytically tractable model is one where the tracer is coupled to bath particles via harmonic interactions \cite{Crisanti2012,Netz2018,Wang2021,Goswami2023}.  While each particle individually follows Markovian dynamics, integrating out the bath degrees of freedom results in the tracer's effective dynamics exhibiting an exponential memory term, which reflects the viscoelastic properties of the environment \cite{Ferrer2021,Guevara-Valadez2023}. Moreover, when the tracer and bath particles are not connected to the same thermal reservoir, a net energy flux arises even with this simple coupling, driving the tracer out of equilibrium. This nonequilibrium condition is evident as the memory kernel in the effective dynamics is no longer directly related to the noise statistics through the fluctuation-dissipation relation \cite{Crisanti2012,Wang2021}. When the interaction is turned off, the system reverts to the equilibrium description of individual particles. Variants of this model have been extensively invoked in recent years to explore the dynamics in granular materials and viscoelastic media \cite{Caprini2024,Ginot2022,Das2023}, as well as in studies of heat engines \cite{Filliger2007,Berut2016} and other complex processes \cite{Medina2018,Wang2021}.

With the above motivations in mind, we consider a system where a passive tracer is linearly coupled to bath particles that are in contact with a different heat bath, creating a nonequilibrium, viscoelastic environment. In such a setting, it is interesting to explore how the tracer searches for a target, especially given the importance of search processes in biochemical reactions, where a reactant must locate a target to initiate a reaction. Search processes are essential in various biological systems, encompassing phenomena such as proteins locating promoter sites on DNA during gene expression \cite{Hippel1989,Bauer2012}, neurotransmitter binding to gated ion channels \cite{Li2009,Northrup1982}, foraging behavior in   animals \cite{Fauchald2003,Monasterio2011}, and visual search and pattern recognition in psychology \cite{Noton19711,Noton19712,Henderson2003}. In a simple random walk within an unbounded domain, a particle typically takes an infinite amount of time to find a target, failing to reach equilibrium or a steady state \cite{Redner2001}.

However, over the past decade, search strategies under stochastic resetting have garnered significant attention \cite{Evans2020}. In stochastic resetting, a diffusing tracer intermittently returns to a specified location at a fixed rate. This mechanism confines the particle, leading to a nonequilibrium steady state and optimizing the search process at specific resetting rates \cite{Evans2011,Friedman2020}. Various extensions of this concept have been explored, including space-dependent resetting rates \cite{Evans2011j}, externally switched confining potentials \cite{Gupta2020}, search processes involving continuous random walks \cite{Méndez2021}, Lévy flights \cite{Kusmierz2014}, fractional Brownian motion \cite{Majumdar2018}, geometric Brownian motion \cite{Vinod2022}, resetting of run-and-tumble particles \cite{Evans2018}, activity-driven diffusions \cite{goswami2021,abdoli2021}, and active fluctuations of membranes \cite{Gupta2014,Singha2023}. Few studies have explored resetting in heterogeneous media with time-dependent diffusion coefficients \cite{Bodrova2019,Bodrova20191}. To the best of our knowledge, resetting in a nonequilibrium bath, as described in this study, has not been extensively investigated. This research aims to address this important gap in the existing literature. We employ numerical simulations combined with standard analytical technique to explore the effects of resetting on tracer dynamics at different relaxation timescales.

In Sec. \ref{sec2} and Sec. \ref{sec3}, we  present the model  and study the dynamical properties of the tracer subjected to stochastic resetting in an unbounded domain, extending our analysis to the relaxation process as the system approaches a steady state. In  Sec. \ref{sec4}, we examine the first-passage properties in the presence of resetting. Section \ref{sec5} explores the effects of resetting when the particle is subjected to an external force, with a particular emphasis on determining the first-passage properties in the presence of a target. We conclude our findings in Sec. \ref{conclusion}. Additional details and full derivations are provided in the Appendices. In Appendices \ref{appen1} and \ref{appen2}, we analytically derive the probability density function (PDF) for both free and forced cases, and discuss the relaxation dynamics in free space. First-passage calculations for the two cases are presented in Appendix \ref{appen3}. An important Fourier and Laplace transform of a function, used to find the propagator in the aforementioned appendices, is provided in Appendix \ref{appen4}. Detailed simulation method is discussed in Appendix \ref{appen5}.

\section{Model \label{sec2}}
Consider a Brownian tracer particle diffusing in a bath at temperature $T_x$, with its position $x$ elastically coupled to the degrees of freedom $y_i$ of $N$ bath particles. These bath particles are connected to a separate reservoir at temperature $T_y.$ Each particle follows Markovian dynamics individually, obeying detailed balance and the Einstein relation, $T_x=\gamma_x D_x$ and $T_y=\gamma_y D_y$ , where $D_x$ and $D_y$ are the diffusivities, and $\gamma_x$ and $\gamma_y$ are the friction coefficients for the tracer and bath particles, respectively. However, due to the coupling between these particles at different temperatures, the environment becomes nonequilibrium, and the dynamics of the tracer exhibits non-Markovianity when the degrees of freedom $y_i$ are integrated out \cite{Ferrer2021,Guevara-Valadez2023}. Instead of explicitly modeling the non-Markovian dynamics, the system is described by two coupled stochastic equations for $x$ and auxiliary variables $y_i.$  Therefore, in the overdamped limit, the Langevin equations for the tracer and the bath particles can be expressed as  
\cite{Ginot2022,Goswami2023}
\begin{subequations}
\begin{align}
\dot{x}(t)=&-\sum_{i=1}^{N}\,k_b \Big(x(t)-y_i(t)\Big)+\eta_x(t),\label{langevin-tracer}\\
\dot{y}_i(t)=&-\gamma k_b\Big(y_i(t)-x(t)\Big)+\eta_y(t),\label{langevin-bath}
\end{align}
\end{subequations}
For simplicity, we consider $N=1$ in this paper. Here, $\eta_x(t)$ and $\eta_y(t)$ are mean-zero Gaussian noises with strengths $D_x$ and $D_y$, respectively, i.e.,
\begin{subequations}
\begin{align}
 \langle\eta_i(t)\rangle &=0,\nonumber\\
 \langle\eta_i(t)\eta_j(t')\rangle &=2D_i\delta(t-t') \delta_{ij},
  \end{align}  
\end{subequations}
where $i,\,j \in \{x,y\}$, $\gamma$ is the ratio of drag coefficients, namely $\gamma=\gamma_x/\gamma_y.$ Here, $k_b$ signifies the interaction (or coupling) strength between the tracer and bath particles, thus dictating a stress-relaxation timescale defined as $\tau_r=k_b^{-1}$. The limiting case with  $k_b \rightarrow 0$  corresponds to the free tracer motion. In the opposite limit, $k_b \rightarrow \infty,$ the system reaches a steady state or an ``effective" equilibrium at a temperature determined by both diffusivities $D_x$ and $D_y.$ It is to note here  that the equations \ref{langevin-tracer} and \ref{langevin-bath} are analogous to two particles in a dimer connected by harmonic springs \cite{Blossey2019}.

\begin{figure*}[htp]
    \centering
    \begin{subfigure}[b]{0.495\textwidth}
        \centering
        \includegraphics[width=\textwidth]{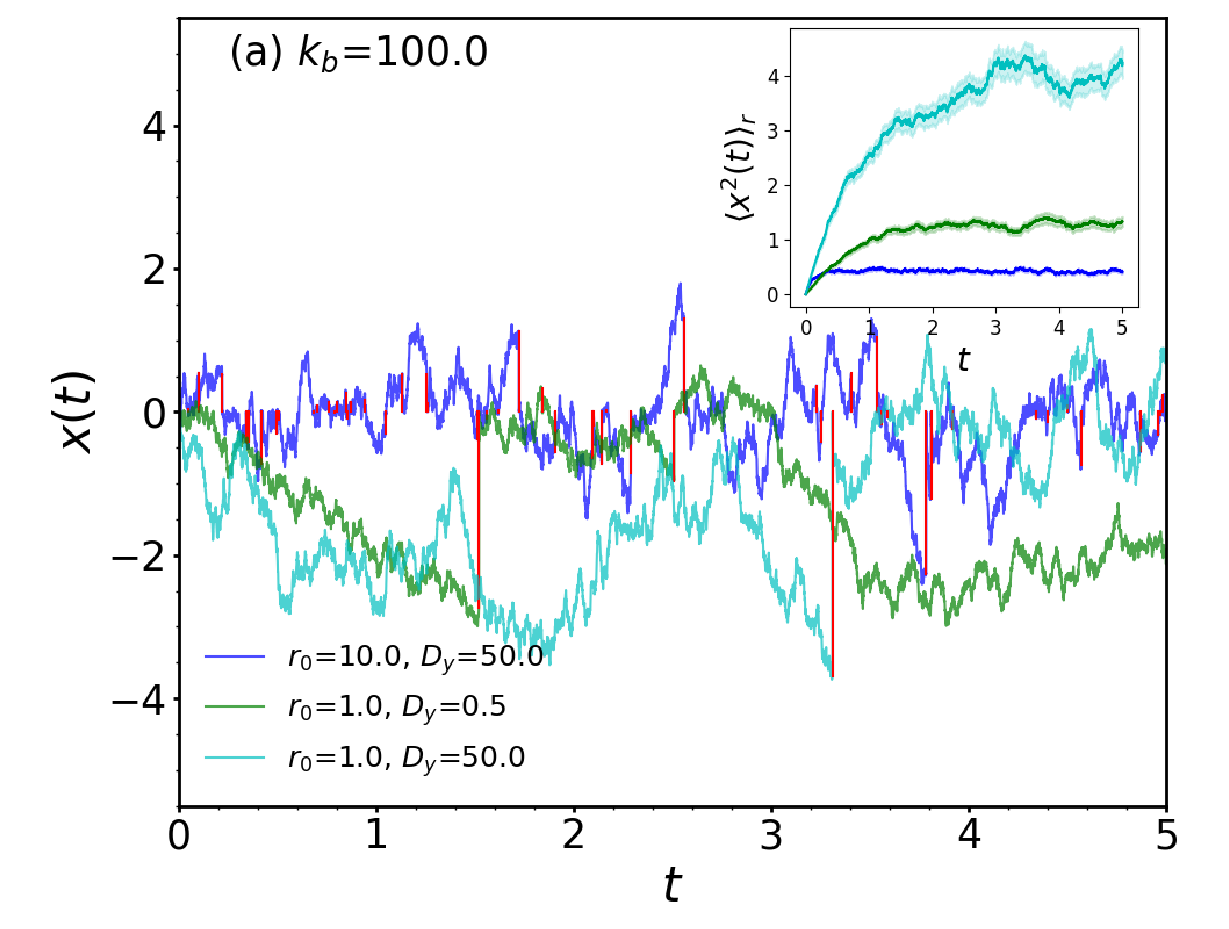}
    \end{subfigure}
    \hfill
    \begin{subfigure}[b]{0.495\textwidth}
        \centering
        \includegraphics[width=\textwidth]{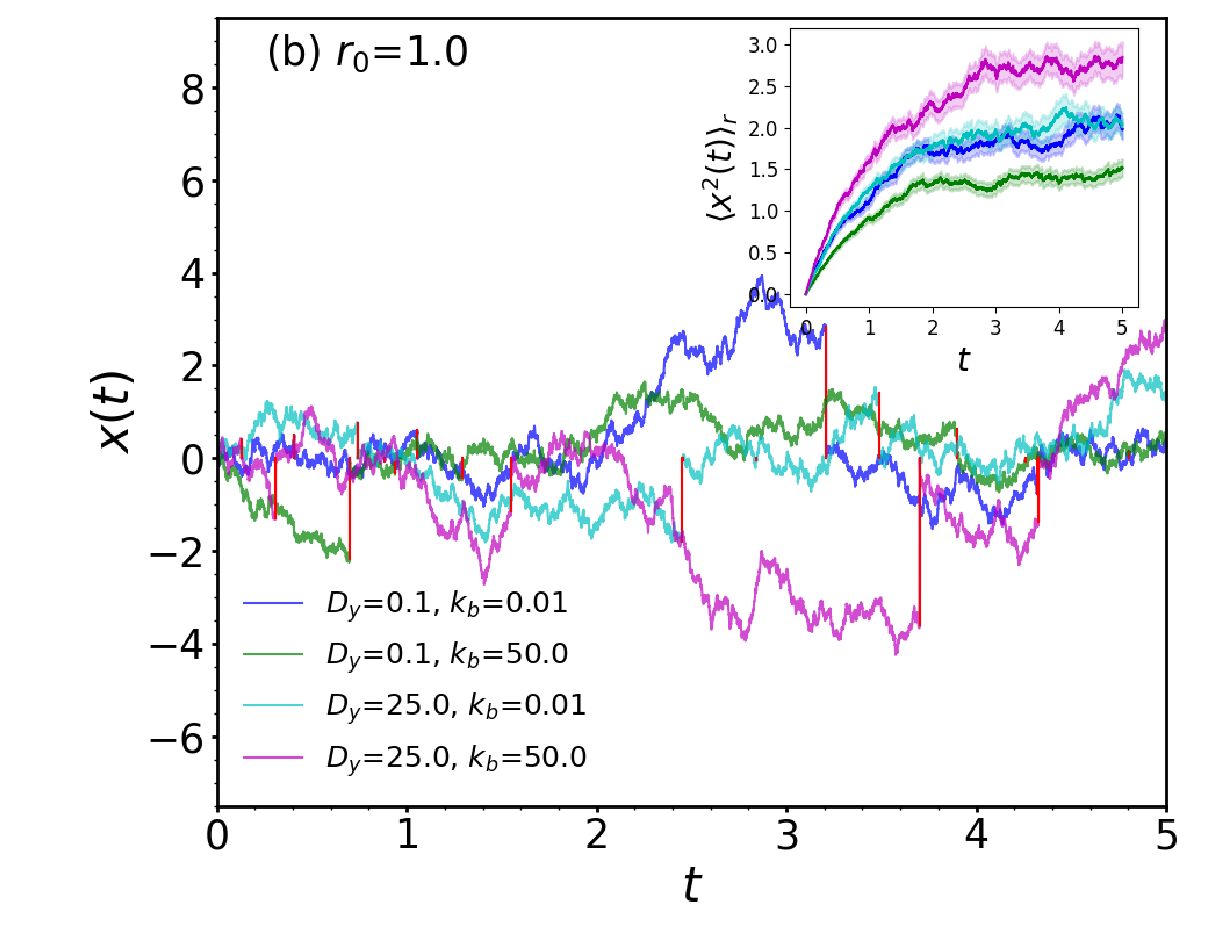}
    \end{subfigure}
\caption{Simulated trajectories of the tracer $x$ over time $t$, where $x$ is reset to its initial position $x(0)=0$ at random times, indicated by vertical red lines, while interacting with a bath particle $y$. Panel (a) presents trajectories of $x$ for two distinct diffusivities of the bath particle at different resetting rates, $r_0.$ Panel (b) illustrates various trajectories of $x$ at a specific resetting rate, $r_0=1.0$, where the tracer interacts with bath particles of either higher or lower diffusivities at two distinct interaction strengths, $k_b$. Insets: The ensemble-averaged MSD values, calculated over $10^3$ trajectories for each parameter set, are plotted as a function of time. The shaded regions indicate the standard error, providing an estimate of the variability around the average MSD.
} 
\label{fig:trajectory}
\end{figure*}

Now consider a situation where the diffusing tracer initially starts at position $x(0)=0$ and is subject to a resetting protocol as follows: the tracer is reset to its initial position $x(0)$ following Poisson statistics with a Poisson rate $r_0$, whereas the variable $y$  is not subject to reset. Therefore, under resetting of $x$, the values of  $x$ and $y$ are updated according to the following stochastic equations 
\begin{subequations}
\begin{align}
x(t+dt)=& x(0),\;\text{with probability } r_0 dt,\label{langevin-tracer_reset1}\\
x(t+dt)=&x(t)-k_b \Big(x(t)-y(t)\Big)dt+\eta_x(t) dt,\nonumber\\
&\qquad\qquad\;\text{with probability } (1-r_0 dt),\label{langevin-tracer_reset2}\\
y(t+dt)=&y(t)-\gamma k_b\Big(y(t)-x(t)\Big)dt+\eta_y(t)dt.\label{langevin-bath_reset}
\end{align}
\end{subequations}
The typical trajectories of the tracer $x(t)$ are depicted in Fig. \ref{fig:trajectory}. Panel (a) shows that as the resetting rate $r_0$ increases, the tracer undergoes more frequent resetting events, as expected. In panel (b), representative trajectories of the tracer are shown for four scenarios, illustrating interactions that are either strong $(k_b > 1)$ or weak $(k_b < 1)$ with bath particles having either lower or higher diffusivity.  Interestingly, the corresponding mean square displacement (MSD) plots reveal that the impact of stronger interactions on the tracer's displacement varies based on the diffusivity of the bath particle. To further explore this behavior, we analyze the dynamical and first-passage properties in the subsequent sections.

\section{Dynamical properties  \label{sec3}}
In the presence of resetting, the propagator for both $x$ and $y$ can be expressed in terms of the last  renewal equation as follows
\begin{align}
P_r(x,y,t|& x(0)=0,y(0)=y_0,0) = e^{-r_0 t} P_0(x,y,t|0,y_0,0)\nonumber\\
& + r_0 \int_{0}^{t}d\tau e^{-r_0 \tau}\,P_0(x,y,\tau|0,y_0,0),
\end{align}
where $P_0$ is the propagator in the absence of resetting. Thus, one can obtain the joint probability density function of $x$ and $y$ after averaging over the initial distribution of $y$,  denoted as $P(y_0)$, where  $P(y_0)=\sqrt{\frac{\gamma k_b}{2\pi D_y}} \exp\left(-\frac{\gamma k_b}{2D_y}y_0^2\right).$ We are particularly interested in the dynamical behavior of the tracer, which can be described by its PDF of $x$, or the marginal distribution of $x$. This can be computed as
\begin{align}
& P_r(x,t| x_0,0)\nonumber\\
&=\int dy\int dy_0 P(y_0)\, P_r(x,y,t| x_0,y_0,0)\nonumber\\
&= e^{-r_0 t} \langle P_0(x,y,t|x_0,y_0,0)\rangle_{y,y_0}\nonumber\\
& + r_0 \int_{0}^{t}d\tau e^{-r_0 \tau}\,\langle P_0(x,y,\tau|x_0,y_0,0)\rangle_{y,y_0},\nonumber\\
&= e^{-r_0 t} P_0(x,t|x_0,0)+ r_0 \int_{0}^{t}d\tau e^{-r_0 \tau}\,P_0(x,\tau|x_0,0),\label{last_renewal_equation}
\end{align}
where $\langle\cdots \rangle_{y,y_0}$ denotes  the averaging over all initial positions $y_0$ and all possible trajectories of $y$. In the last line of Eq. (\ref{last_renewal_equation}), the first term on the right-hand side (RHS) represents the scenario where no reset event has occurred up to time $t$. The second term corresponds to the case where a reset occurs at time $t-\tau$ and no subsequent reset occurs thereafter. 
 In the above equation, the reset-free propagator can be expressed as  [see Eq. (\ref{pdf_px4})]
\begin{align}
 P_0(x_f,t|x_i,0)=\sqrt{\frac{1}{4\pi\mathcal{C}_0^2(t)}} \exp\left(-\frac{\left(x_f- x_i \mathcal{S}(t)\right)^2}{4 \mathcal{C}_0^2(t)}\right), \label{pdf_px4_0}
\end{align}
where  
\begin{subequations}
\begin{align}
&\mathcal{C}_0^2(t)=a_1+a_2 t + a_{11} e^{-k_b(1+\gamma)t} +a_{12} e^{-2k_b(1+\gamma)t},\label{C_0^2}\\
&\mathcal{S}(t)= \frac{\gamma+e^{-k_b(1+\gamma)t}}{\gamma+1} \label{S0},
\end{align}
\end{subequations}
and the other parameters  involved in the above expressions are as follows 
\begin{subequations}
\begin{align}
& a_1=\frac{D_y (1-2 \gamma) + D_x \gamma (1+4\gamma)}{2 \gamma k_b (\gamma + 1)^3}, \label{a1}\\
& a_2=\frac{D_x \gamma^2 + D_y }{(\gamma + 1)^2},\label{a2}\\
& a_{11}=\frac{\left[2D_y (\gamma-1)-4 D_x \gamma^2 \right]}{2 \gamma k_b (\gamma + 1)^3},\label{a11}\\
& a_{12}=\frac{ D_y -D_x \gamma }{2 \gamma k_b (\gamma + 1)^3}\label{a12}.
\end{align}    
\end{subequations}
The full derivation of the propagator and other supplementary information are relegated to Appendix \ref{appen1}, while only the essential results are presented in the main text for the convenience of readers and to maintain the flow of the paper.
\begin{figure}
\centering
\includegraphics[width=1\linewidth]{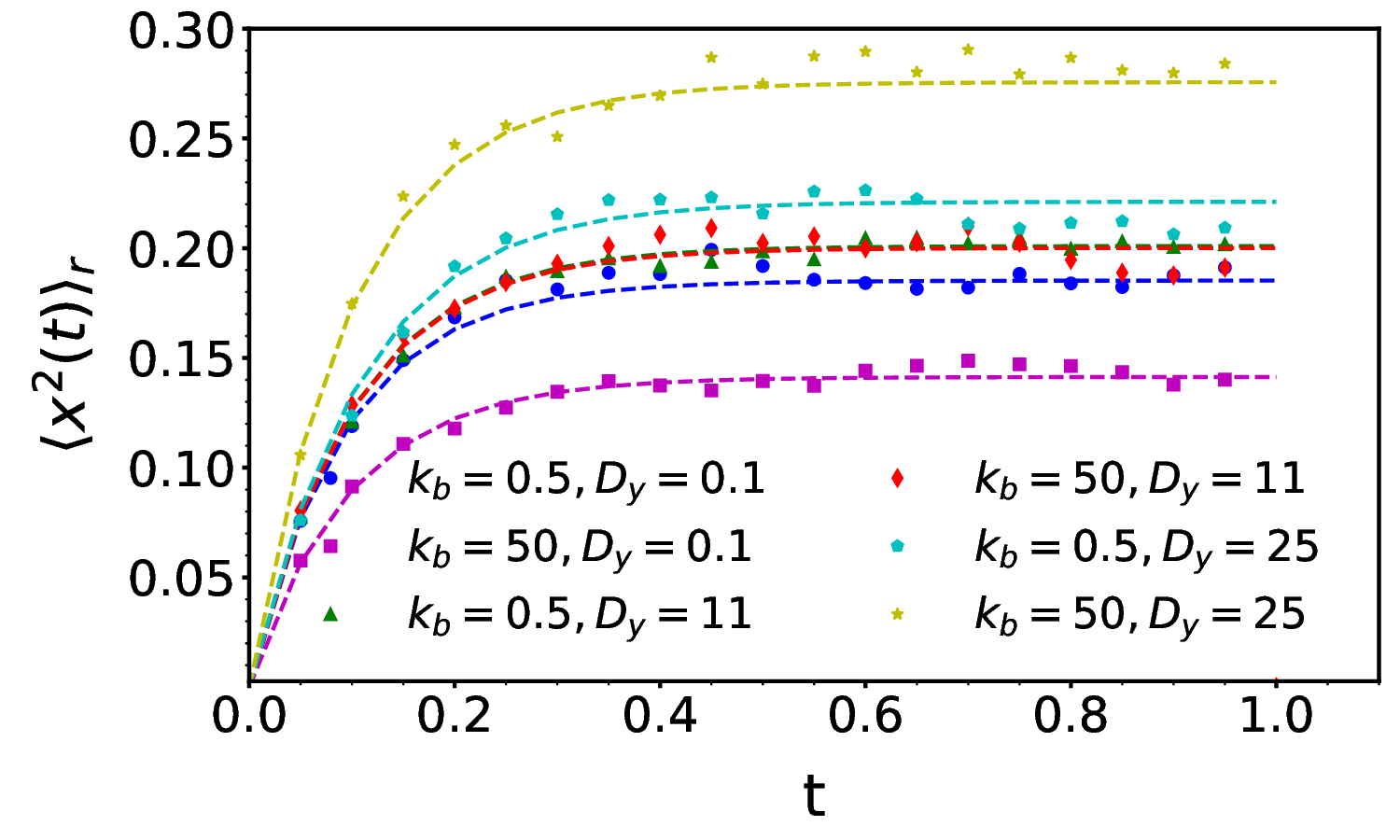}
\caption{Plot of the MSD as a function of time $t$ for various values of diffusivity $D_y$ and  interaction strength $k_b.$ The simulation data  are represented by symbols, while the dashed lines depict analytical results obtained from Eq. \ref{msd_exact_0}. The resetting rate is set to $r_0=10$, and  $D_y=11.0$ corresponds to $D_y^c$. Additional parameters and simulation details are provided in Appendix \ref{appen5}. } \label{fig:msd}
\end{figure}

\begin{figure}
\centering
\includegraphics[width=1\linewidth]{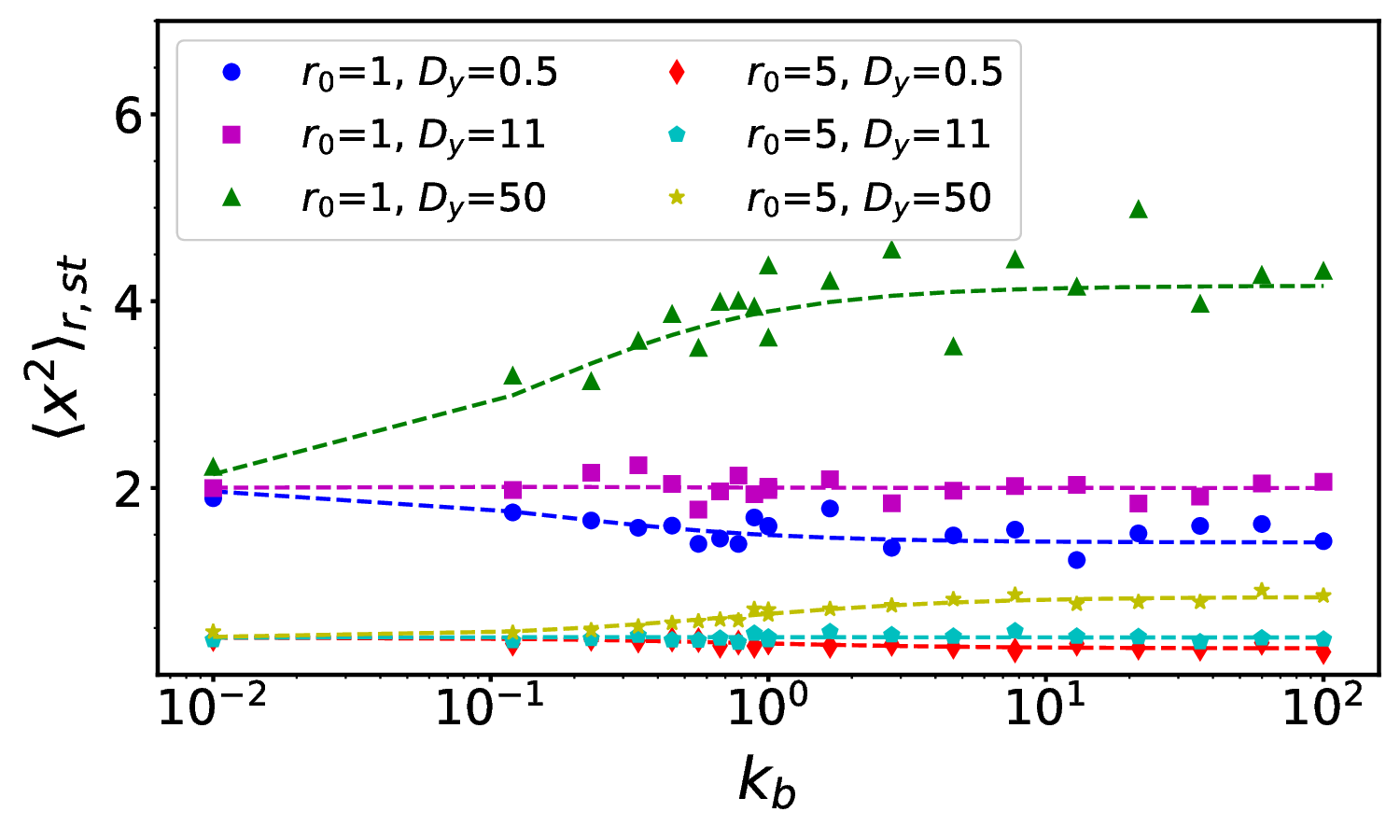}
\caption{Steady-state value of the MSD is plotted as a function of $k_b$ for two different $r_0$ values in a linear-log scale. The analytical expression, as given in Eq. (\ref{msd_longt_0}), is represented by dashed lines and shows good agreement with the simulation data depicted by symbols. Here, $D_y=11.0$ corresponds to $D_y^c.$  Other details are provided in Appendix \ref{appen5}.} \label{fig:msd_kb}
\end{figure}

 \begin{figure*}[htp]
    \centering
    \begin{subfigure}[b]{0.45\textwidth}
        \centering
        \includegraphics[width=\textwidth]{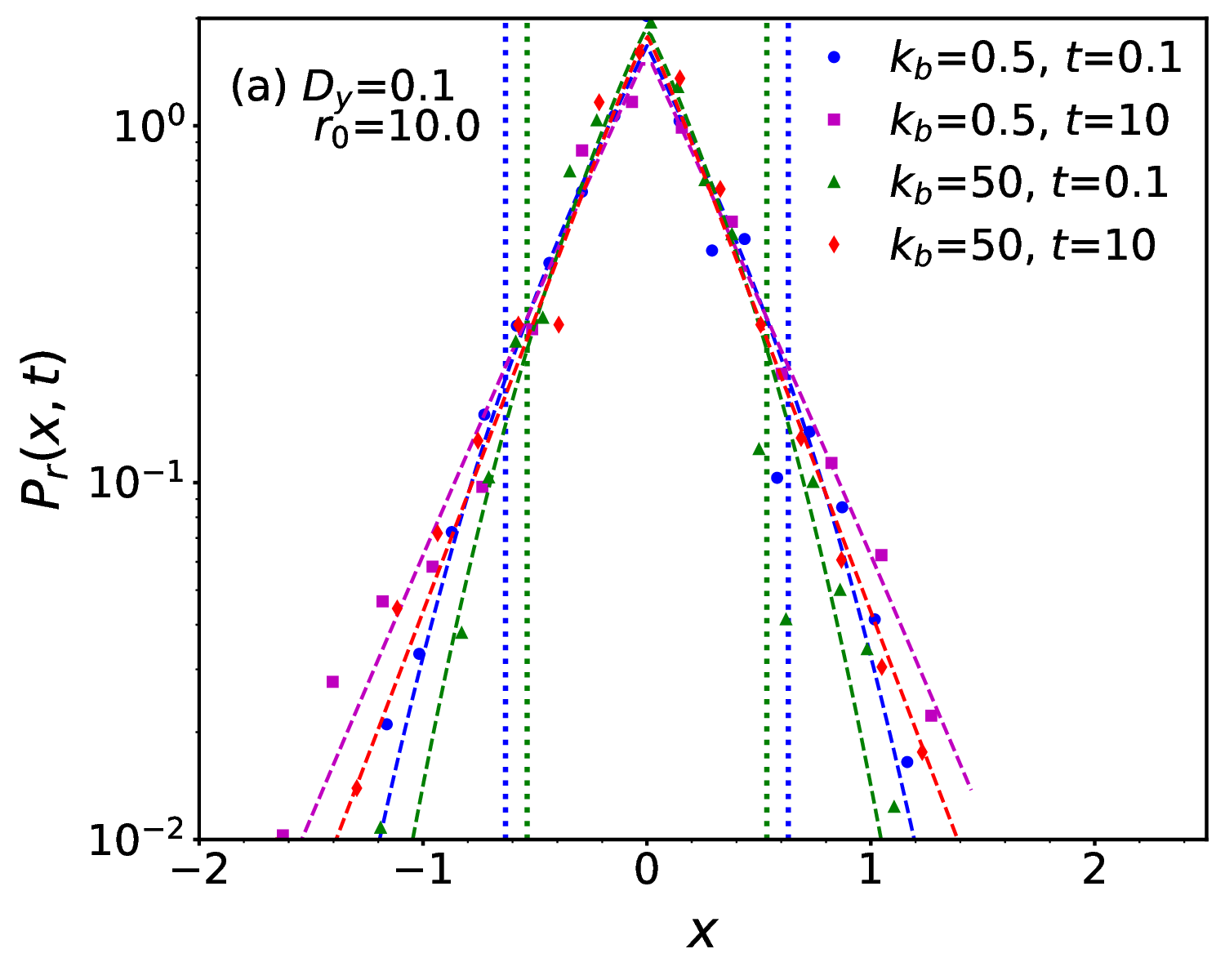}
    \end{subfigure}
    \hfill
    \begin{subfigure}[b]{0.45\textwidth}
        \centering
        \includegraphics[width=\textwidth]{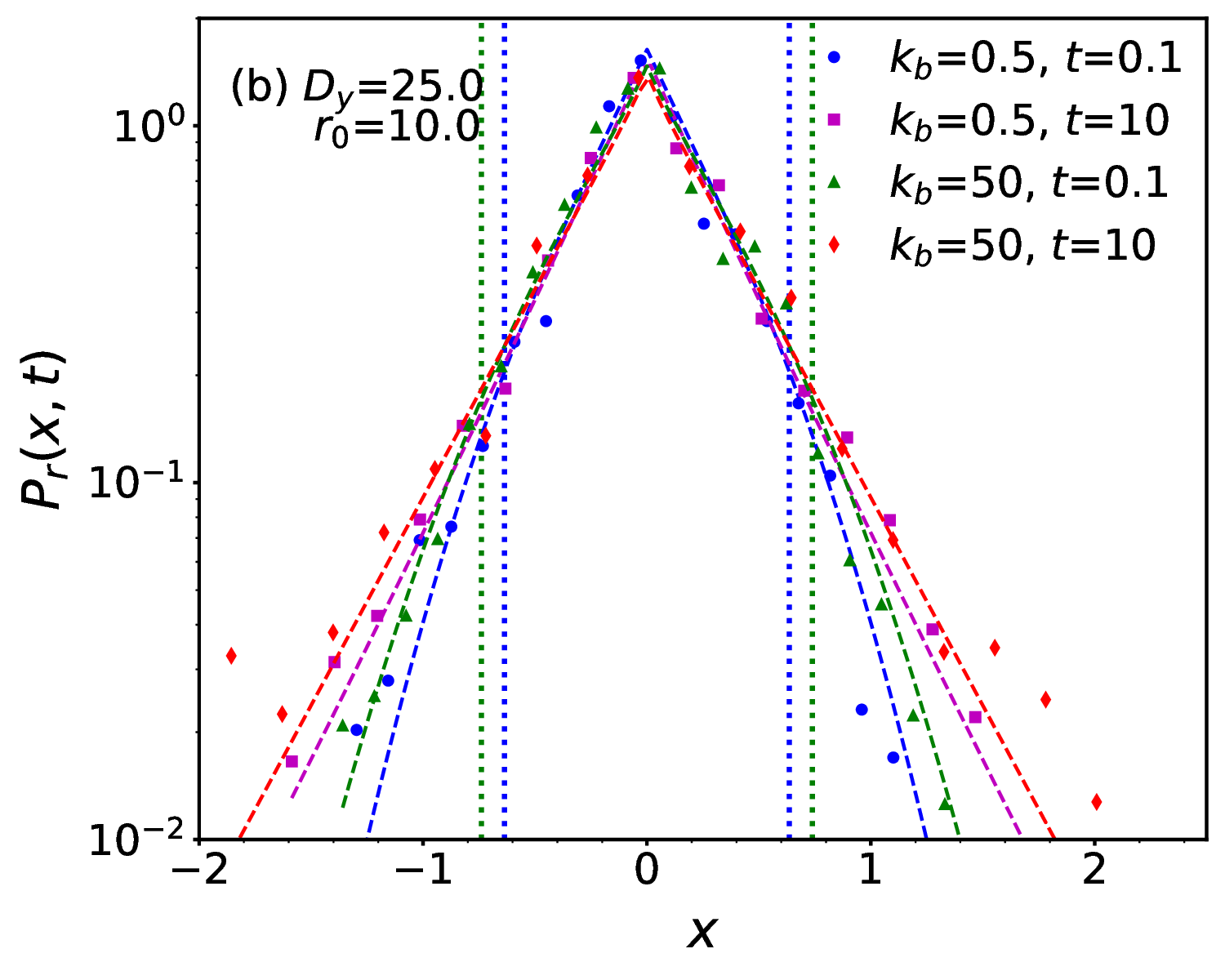}
    \end{subfigure}
\caption{ Logarithmic plot of the PDF as a function of position $x$ at different times for two values of $k_b.$ The simulation data, represented by symbols, are compared with the analytical results computed numerically from Eq. (\ref{last_renewal_equation}), depicted by lines. The two  vertical  dotted lines indicate the positions $x^*,$  marking the boundaries between the transient and steady-state regimes of the distribution at a particular time $t=0.1$. Here,  $r_0=10.0$ and $D_y^c=11.0,$ with other parameters as  in Fig. \ref{fig:msd}.} \label{fig:pdf_reset}
\end{figure*}

The second moment of the PDF, or the MSD of the tracer, can be computed using Eq. (\ref{last_renewal_equation}) as
\begin{align}
\langle x^2(t)\rangle_r =  e^{-r_0 t}\langle x^2(t)\rangle_0+ r_0 \int_{0}^{t}d\tau e^{-r_0 \tau}\,\langle x^2(\tau)\rangle_0 \label{msd},
\end{align}
where $\langle x^2(\tau)\rangle_0$ is the MSD for the reset-free case, which can be calculated from the reset-free propagator given in Eq. (\ref{pdf_px4_0}). In practice,  $\langle x^2(\tau)\rangle_0$ is directly obtained from the Fourier transform of $P_0(x,t|x_0,0),$ as outlined in Eq. (\ref{msd_fourier}). The exact expression for the MSD under resetting is given by [see Eq. (\ref{msd_exact})]
\begin{align}
\langle x^2(t)\rangle_r = 2 e^{-r_0 t} \mathcal{C}_0^2(t) + 2r_0  \mathcal{C}_{r0}^2(t),\label{msd_exact_0}
\end{align}
where 
\begin{align}
\mathcal{C}_{r0}^2(t)&=\frac{a_1(1-e^{-r_0 t})}{r_0}+\frac{a_2 [1-(1+r_0 t)e^{-r_0 t}]}{r_0^2}\nonumber\\
&+\frac{a_{11}}{r_0+k_b(1+\gamma)}(1-e^{-[r_0+k_b(1+\gamma)] t})\nonumber\\
& +\frac{a_{12}}{r_0+2 k_b(1+\gamma)}(1-e^{-[r_0+2 k_b(1+\gamma)] t}). \end{align}
The analytical results are presented in Figs. \ref{fig:msd}-\ref{fig:msd_kb} and compared with the simulation data, demonstrating excellent agreement. In the limit $t\rightarrow 0$, the MSD grows linearly with time, independent of the diffusivity $D_y$ [cf. Eq. (\ref{msd_shortt})], viz. $ \langle x^2(t)\rangle_r \approx  2D_x t.$
In the long-time limit, the MSD eventually converges to a constant value
\begin{align}
    \langle x^2\rangle_{r,st} \approx 2 a_1+\frac{2 a_2}{r_0}+\frac{2r_0 a_{11}}{r_0+k_b(1+\gamma)}+\frac{2 r_0 a_{12}}{r_0+2 k_b(1+\gamma)},\label{msd_longt_0}
\end{align}  indicating that the system reaches a steady state [cf. Eq. (\ref{msd_longt})]. As expected, $\langle x^2\rangle_{r,st}$ decreases with increasing $r_0$, but for the same $r_0$, the MSD clearly depends on the interaction strength $k_b$ as well as $D_y$, as seen in  Fig. \ref{fig:msd_kb}. To examine the dependence on these parameters, we first consider two limiting cases for $k_b$. In the weak-coupling limit ($k_b\rightarrow 0$), we recover the usual result   $\langle x^2\rangle_{r,st} \approx \frac{2 D_x}{r_0},$  corresponding to a thermal bath. 
For large $k_b$ and in a strictly nonequilibrium environment,  the MSD approaches $\langle x^2\rangle_{r,st} \approx \frac{2 a_2}{r_0}$, where $a_2$ is given by Eq. (\ref{a2}). This implies that the MSD increases with larger $D_y$.  
The variation of MSD, $\langle x^2\rangle_{r,st}$, as depicted in Fig. \ref{fig:msd_kb} for intermediate $k_b$ values, highlights three distinct regimes for different values of $D_y$. When $D_y=D_x(2 \gamma+1),$ denoted as $D_y^c$, the MSD becomes $\langle x^2\rangle_{r,st}=\frac{2 D_x}{r_0}$, which is independent of $k_b$. It is important to note that $D_y^c$ does not correspond to the scenario in which all particles, including the tracer and bath particles, are in contact with the same thermal reservoir $(D_y=\gamma D_x)$. In that case, the tracer's effective diffusivity is reduced by a factor of $\frac{k_b\gamma +r_0}{k_b\gamma+k_b +r_0},$ which depends on the coupling strength $k_b$. In this way, the effect of resetting contrasts with the confinement of the tracer in a harmonic potential, where $D_y=\gamma D_x=D_y^{nc}$ was  found \cite{Goswami2023}. We also emphasize that in the absence of resetting, the condition $D_y = D_y^{nc}$ corresponds to a thermal bath, whereas bath particles with diffusivity $D_y > D_y^{nc}$ and $D_y < D_y^{nc}$ are referred to as ``hot" and ``cold" particles, respectively. In the presence of resetting, for $D_y < D_y^c$, $\langle x^2 \rangle_{r,st}$ decreases as the interaction strength $k_b$ increases, while the opposite trend is observed for $D_y > D_y^c$.

The PDF in the presence of resetting is calculated using the free propagator without resetting, given by Eq. (\ref{pdf_px4_0}), through the renewal equation [Eq. (\ref{last_renewal_equation})]. The derivation of the PDF is discussed in detail in Appendix \ref{appen1}.  The tracer initially begins at $x(0)=0$, corresponding to a delta function distribution centered at $x=0$. Over time, as the distribution gradually spreads in space, as illustrated in  Fig. \ref{fig:pdf_reset},  the central part of the PDF relaxes to a time-independent steady state, while the tail of the PDF, extending beyond the crossover position $x^*$, remains transient. This behavior is discussed in further detail in  Appendix \ref{appen2}. 
In viscoelastic media, where tracer motion is retarded---characterized by higher $k_b$ values and low diffusivity of bath particles $(D_y<D_y^c)$---$x^*$ occurs comparatively closer to $x=0$, indicating a slower approach to steady state, as shown in Fig. \ref{fig:pdf_reset} and Fig. \ref{fig:xstar}. Conversely, when  $D_y>D_y^c$, which can describe activity-induced media such as the cytoplasmic environment, stronger interactions  shift $x^*$ toward larger values, implying that the steady state is established over a broader spatial region.
 
\begin{figure*}[htp]
    \centering
    \begin{subfigure}[b]{0.45\textwidth}
        \centering
        \includegraphics[width=\textwidth]{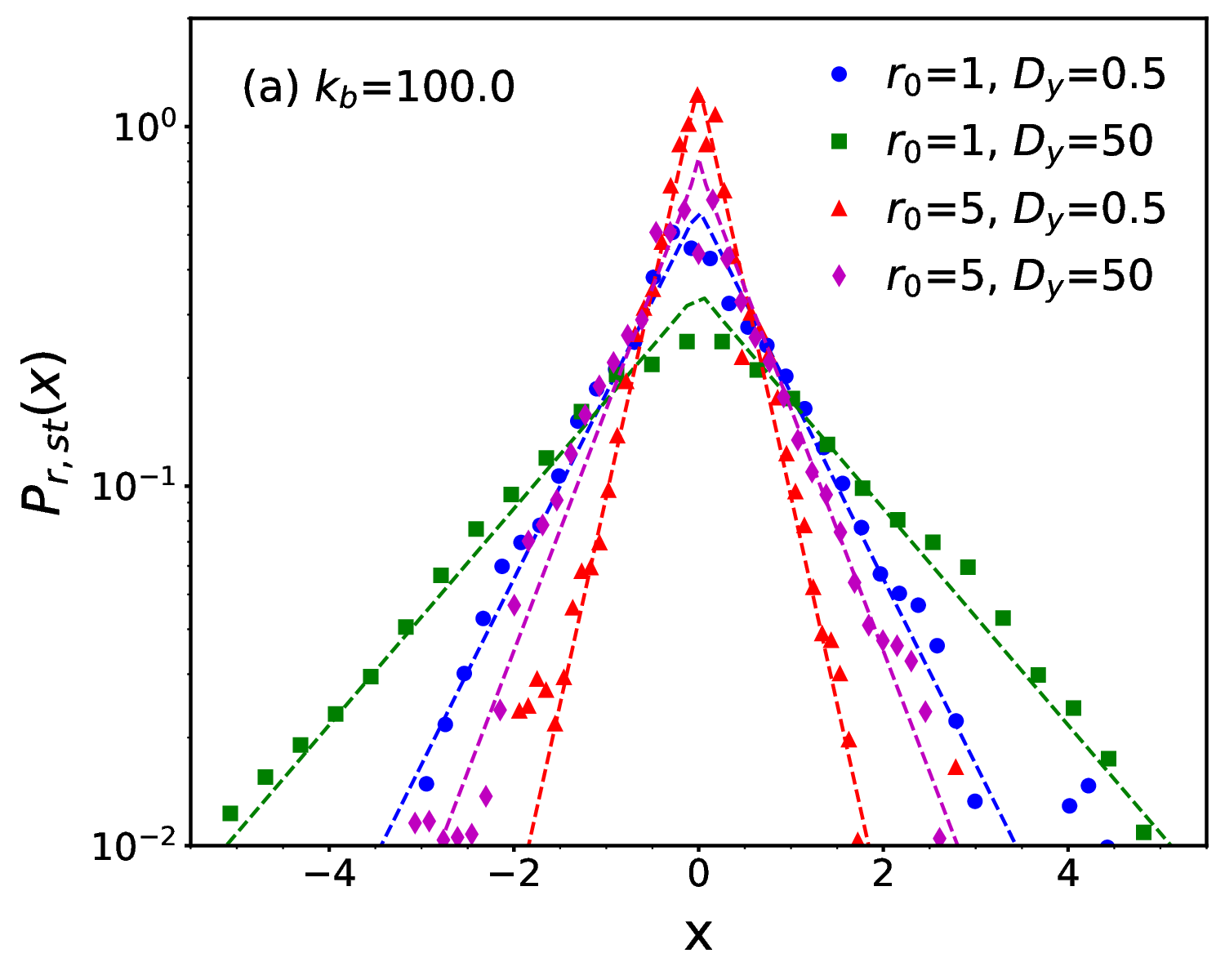}
    \end{subfigure}
    \hfill
    \begin{subfigure}[b]{0.45\textwidth}
        \centering
        \includegraphics[width=\textwidth]{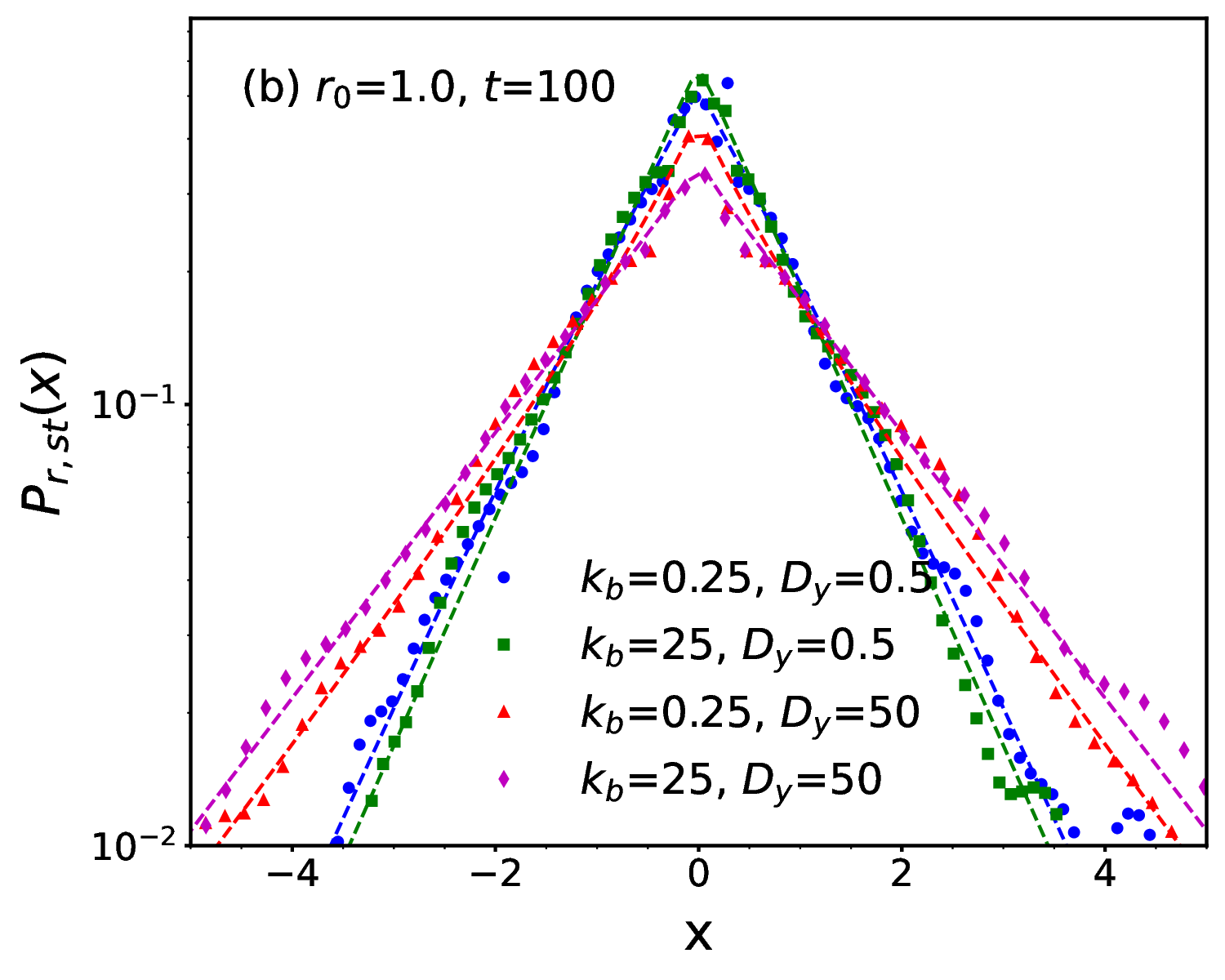}
    \end{subfigure}
\caption{Logarithmic plot of the steady-state PDF: Simulation result (symbols) shows good agreement with the analytical result (dashed line). In panel (a), plots are shown for a higher value of $k_b=100$ across different values of $D_y$, with analytical results given by Eq. (\ref{pdf_reset_st_approx_kbl_0}). Panel (b) illustrates the PDF for various $k_b$ values across different $D_y$ regimes, defined relative to $D_y^c=11.0.$ The analytical predictions, derived from Eq. (\ref{pdf_reset_st}),  are shown for comparison.  Other parameters are as specified in Appendix \ref{appen5}.}
 \label{fig:pdf_reset_kb}
\end{figure*}

Analytically determining the exact steady-state density is not feasible for all cases.  However, in both limits $k_b \ll 1$ and $k_b \gg 1,$  exact results can be obtained [see Eqs. (\ref{pdf_reset_st_approx_kbl})-(\ref{pdf_reset_st_approx_thermal})], as given by
\begin{subequations}
\begin{align}
& P_{r,st}(x) \approx \frac{1}{2} \sqrt{\frac{r_0}{D_x}} e^{-|x| \sqrt{\frac{r_0}{D_x}}},\,\text{for }\;k_b \ll 1, \label{pdf_reset_st_approx_thermal_0}  
\end{align} 
\begin{align}
& P_{r,st}(x) \approx \frac{1}{2} \sqrt{\frac{r_0}{a_2}} e^{-|x| \sqrt{\frac{r_0}{a_2}}} ,\,\text{for }\;k_b \gg 1. \label{pdf_reset_st_approx_kbl_0}
\end{align} 
\end{subequations}
Thus, for very small values of $k_b$, the PDF approximates a Laplace distribution, similar to the one previously reported for a thermal bath \cite{Evans2011}. At large $k_b$ values, the PDF also takes a Laplace form, with the scale parameter now determined by both $D_x$ and $D_y$.
 Based on previous MSD analysis and the PDF, the effective diffusivity of the tracer in a nonequilibrium bath is found to be $a_2$. When $a_2<D_x$  (for $D_y<D_y^c$) or $a_2>D_x$ (for $D_y>D_y^c$), the distribution appears more constricted or broadened, respectively, as illustrated in Fig. \ref{fig:pdf_reset_kb}. Therefore, we refer to a bath particle with diffusivity $D_y<D_y^c$ as a ``cold" particle, while $D_y>D_y^c$ represents a ``hot" particle. Consequently, the tracer's movement is either restricted or enhanced, depending on its strong coupling with cold or hot bath particles, respectively. 
In the strong coupling regime, where all particles are in thermal equilibrium $(D_y=D_y^{nc})$, the tracer dynamics slow down due to a reduction in its  effective diffusivity by the factor $\frac{\gamma}{\gamma+1},$ which  implies that $D_y^{nc}<D_y^c.$

\section{First-Passage properties \label{sec4}}

\begin{figure*}[htp]
    \centering
    \begin{subfigure}[b]{0.46\textwidth}
        \centering
        \includegraphics[width=\textwidth]{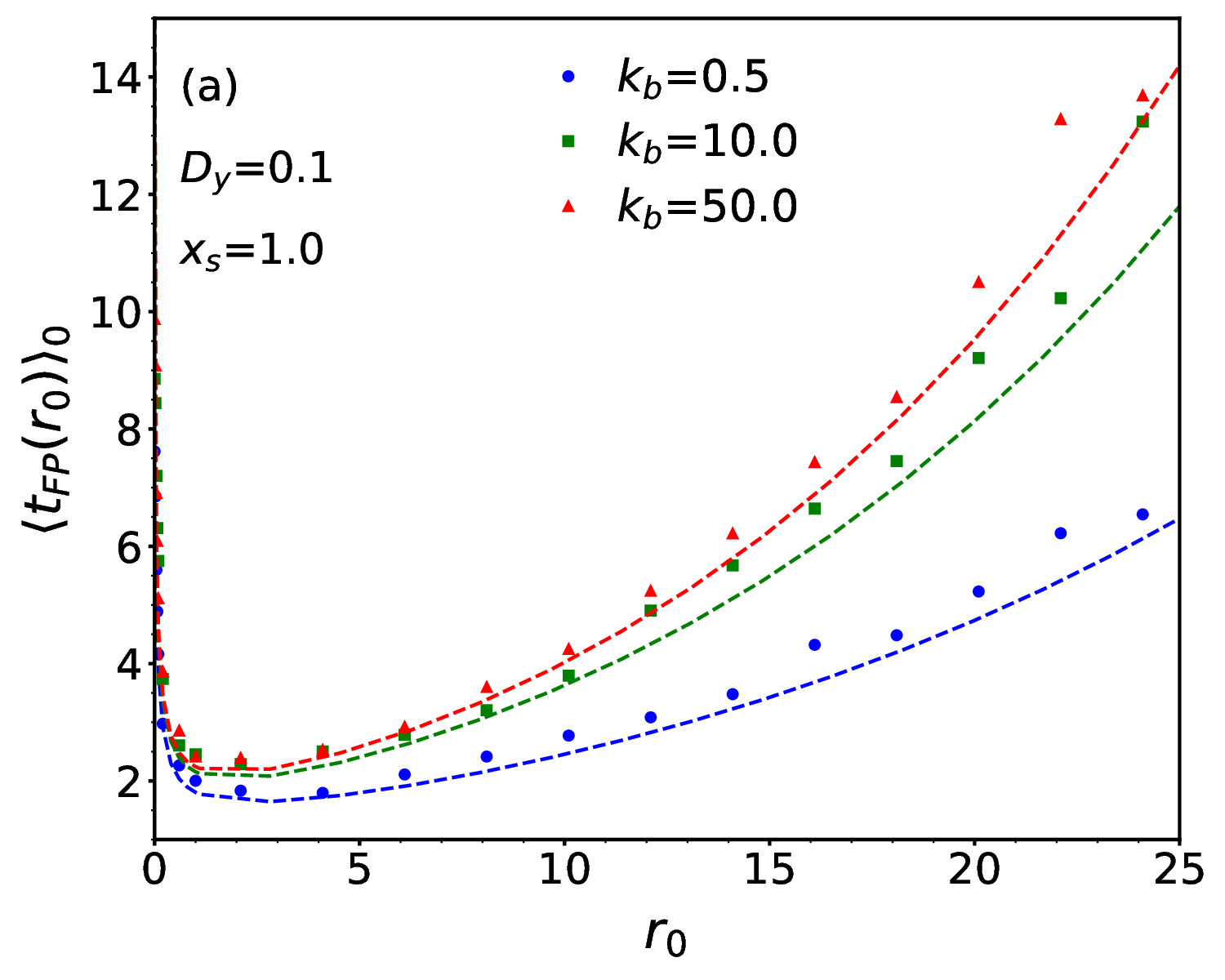}
    \end{subfigure}
    \hfill
    \begin{subfigure}[b]{0.45\textwidth}
        \centering
        \includegraphics[width=\textwidth]{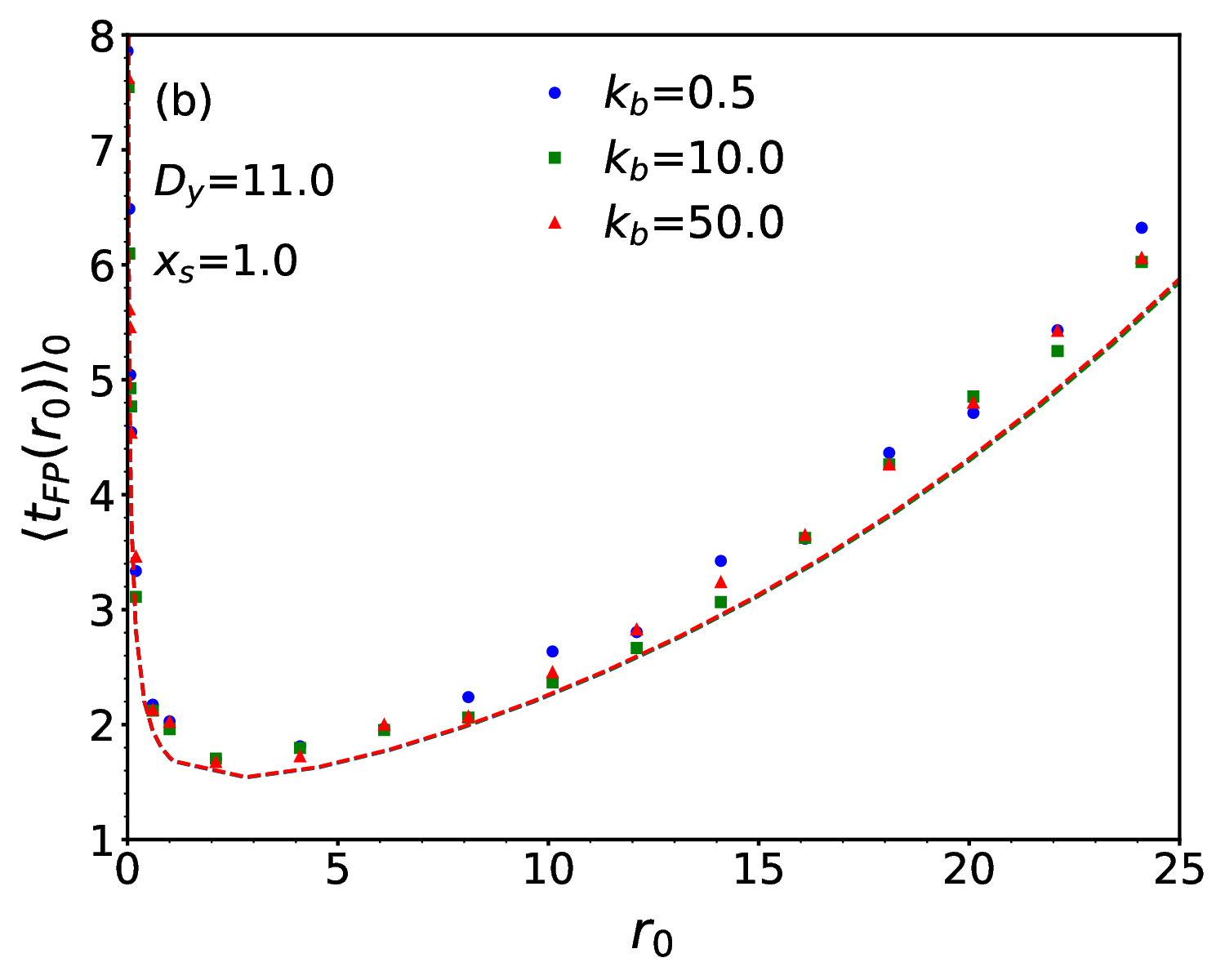}
    \end{subfigure}
        \vspace{0.2cm}
        \centering
    \begin{subfigure}[b]{0.495\textwidth}
        \centering
        \includegraphics[width=\textwidth]{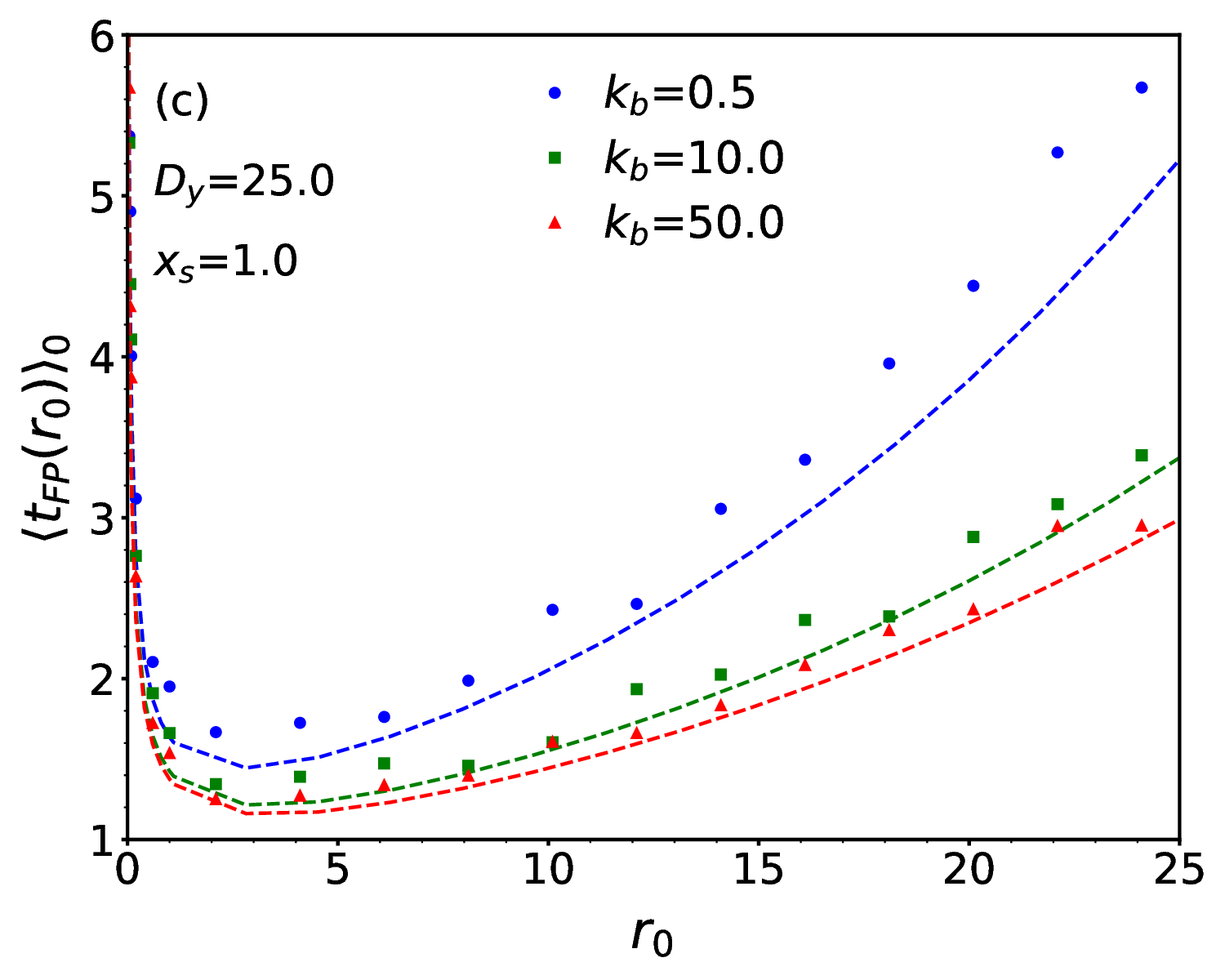}
    \end{subfigure}
\caption{Plot of the MFPT  in the presence of an absorbing target at $x_s=1.0$ as a function of the resetting rate $r_0.$ The results are shown across three panels, representing the three regimes: (a) $D_y < D_y^c$, (b) $D_y = D_y^c=11.0$, and (c) $D_y > D_y^c$.  The simulation results for $\langle t_{FP}(r_0)\rangle_0$, represented by three distinct symbols corresponding to three different values of the coupling strength $k_b$, are compared with the semi-analytical results [Eq. (\ref{tfp_free_0})], depicted by dashed lines. Other parameters are consistent with those in  Fig. \ref{fig:msd}.} \label{fig:meantime_kb_xs}
\end{figure*}

\begin{figure}
\centering
\includegraphics[width=1\linewidth]{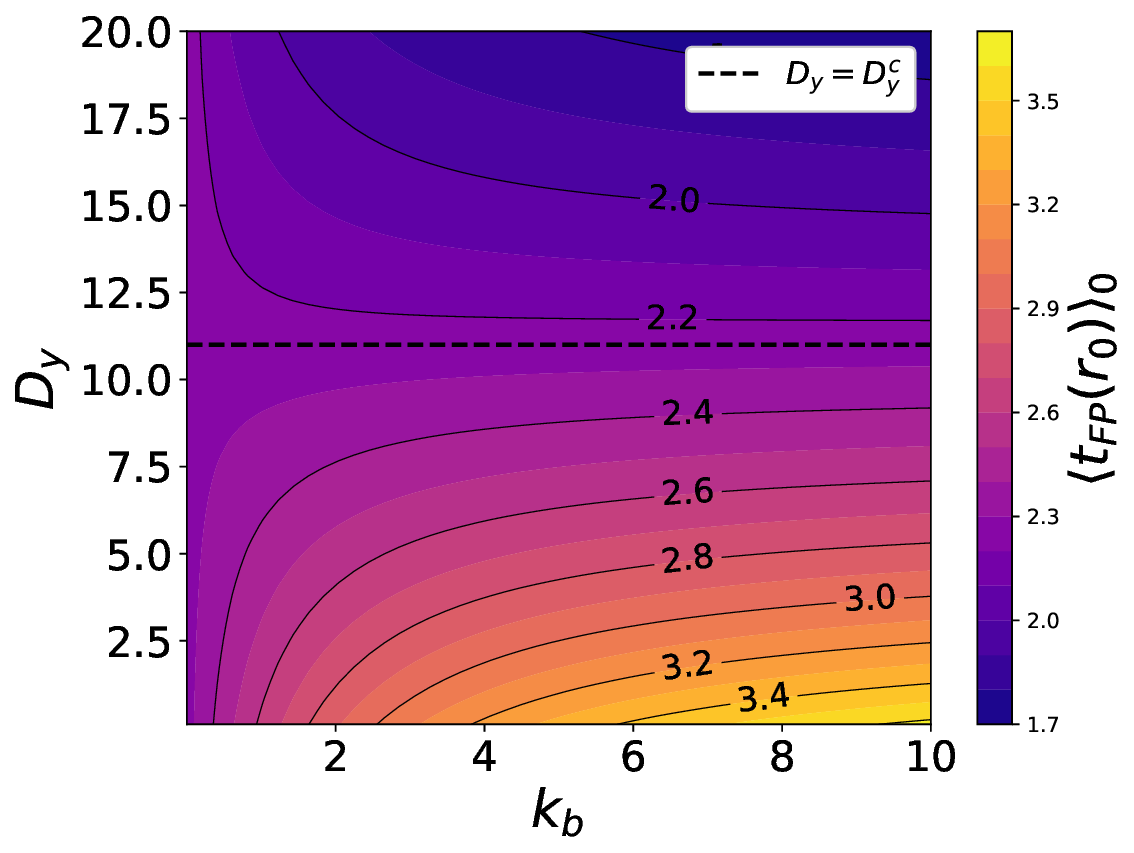}
\caption{Contour plot of the MFPT computed from Eq. (\ref{tfp_free_0}) as a function of $k_b$ and $D_y$ for a resetting rate $r_0=10.0$ and $D_y^c=11.0$ (black dashed line), with parameters $D_x=1$ and $x_s=1.0$. Other parameters are the same as those in   Fig. \ref{fig:meantime_kb_xs}. } \label{fig:meantime_kb_xs_contour}
\end{figure}

Here we are interested in determining the first-passage properties of our process in the presence of a fixed target or a absorbing sink at $x=x_s$ under the same resetting protocol as described earlier. We assume that whenever the tracer encounters the target, it is completely absorbed upon its first visit, and the time is recorded as the first-passage time (FPT). By considering the FPT for all trajectories, we derive the mean first-passage time (MFPT), given by [see Appendix. \ref{appen3}]
\begin{align}
\langle t_{FP}(r_0)\rangle_0=\frac{1}{r_0}\Bigl(\frac{ \widetilde{P}_0(0,r_0|0,0)}{ \widetilde{P}_0(x_s,r_0|0,0)}-1\Bigr),\label{tfp_free_0}
\end{align}
where $\widetilde{P}_0(x_s,r_0|x_0,0)=\int_{0}^{\infty}d\tau\, e^{-r_0 \tau}\,P_0(x,\tau|x_0,0).$ 
The semi-analytical results derived from the above equation align well with simulation data, showing a consistent trend across both methods,  as illustrated  in  Fig. \ref{fig:meantime_kb_xs}. We observe that under resetting, the MFPT remains finite between the two limits, $r_0 \ll 1$ and $r_0 \gg 1$, and varies non-monotonically as a function of $r_0$, a behavior commonly reported in thermal bath \cite{Evans2020}. Typically, there exists an optimal resetting rate, denoted as $r_0^*$, where the MFPT is minimized  for each combination of $k_b$ and $D_y$.

As shown in panels (a)-(c) of Fig. \ref{fig:meantime_kb_xs}, the MFPT exhibits contrasting trends with varying interaction strength $k_b$ at any finite resetting rate $r_0$ across three distinct regimes: Regime I ($D_y < D_y^c$), Regime II ($D_y = D_y^c=D_x[2 \gamma+1]$), and Regime III ($D_y > D_y^c$). In Regime II, where $D_y = D_y^c$, MFPT curves for all $k_b$ values overlap. The three regimes are further highlighted in Fig. \ref{fig:meantime_kb_xs_contour}, which presents a contour plot of the MFPT in the $k_b$–$D_y$ plane for a specific resetting rate. The contour plot reveals that across the $D_y = D_y^c$ line, the MFPT increases with $k_b$ in Regime I, decreases with $k_b$ in Regime III, and remains largely unaffected in Regime II. These observations align with our earlier analysis of  dynamical properties of free tracer, leading to the conclusion that strong interactions facilitate the search process when the tracer is coupled with a hot bath particle ($D_y > D_y^c$) while hindering it when coupled with a cold particle ($D_y < D_y^c$).

Only in the limits $k_b\ll 1$ and $k_b\gg 1$, the Laplace transform of the reset-free propagator can be determined from Eqs. (\ref{pdf_reset_st_approx_thermal_0})-(\ref{pdf_reset_st_approx_kbl_0}). Plugging these into Eq. (\ref{tfp_free_0}), one obtains the exact MFPT as given in Eqs. (\ref{tfp_free_approx_0}) and (\ref{tfp_free_approx}). For ease of reference, the MFPT expressions in both limits are reiterated below, viz.,
\begin{align}
 \langle t_{FP}(r_0)\rangle_0 \approx   
 &\begin{cases}
 & \frac{1}{r_0}\left[e^{x_s\sqrt{\frac{r_0}{D_x}}}-1\right],\,k_b \ll 1\nonumber\\
& \frac{1}{r_0}\left[e^{x_s\sqrt{\frac{r_0}{a_2}}}-1\right],\,k_b \gg 1. 
\end{cases}
\\ \label{tfp_free_approx_0} 
\end{align}
The above equation, along with its approximation in Eq. (\ref{T10}), reinforces the following key points:
 
 i. Based on the values of $a_2$, three distinct regimes emerge, with regimes I and III separated by regime II. The regime II is defined by the condition $a_2=D_x$, or equivalently, when $D_y=D_y^c=D_x(2\gamma+1)$, as elaborated above.
 
 ii. Across the three regimes, one can see that $\langle t_{FP}(r_0)\rangle_0$ in the strong coupling limit follows the order: $\langle t_{FP}(r_0)\rangle_0 \text{ in regime I} >\langle t_{FP}(r_0)\rangle_0 \text{ in regime II}>\langle t_{FP}(r_0)\rangle_0 \text{ in regime III},$ consistent with $a_2$ values such that $a_2 \text{ in regime I} <D_x \text{ in regime II}<a_2 \text{ in regime III}$ [cf. Fig. \ref{fig:rstar_kb}(b)].
 
 iii. As expected, the MFPT decreases when the target is positioned closer to the tracer's initial location, governed by the exponential dependence on the target position $x_s$, as also illustrated in Fig. \ref{fig:meantime_kb_xs1}.
 
The variation of the optimal resetting rate $r_0^*$ and the behavior of MFPT at $r_0^*$ are illustrated in Fig. \ref{fig:rstar_kb}. As mentioned earlier, the tracer coupled to a hot particle can diffuse further from the target due to exploration over a larger spatial region, necessitating more frequent resetting to redirect it toward the target, thereby increasing $r_0^*$. In contrast, for a tracer coupled to a cold bath particle,  $r_0^*$ generally decreases with increasing $k_b$, as stronger interactions confine the tracer and tend to keep it near $x=0$, as shown in panel (a) of Fig. \ref{fig:rstar_kb}. Although  $r_0^*$ exhibits slight non-monotonic behavior with respect to $k_b$, it predominantly adheres to this trend.

\section{Resetting under constant external force $F_0$ \label{sec5}}

\begin{figure*}[htp]
    \centering
    \begin{subfigure}[b]{0.495\textwidth}
        \centering
        \includegraphics[width=\textwidth]{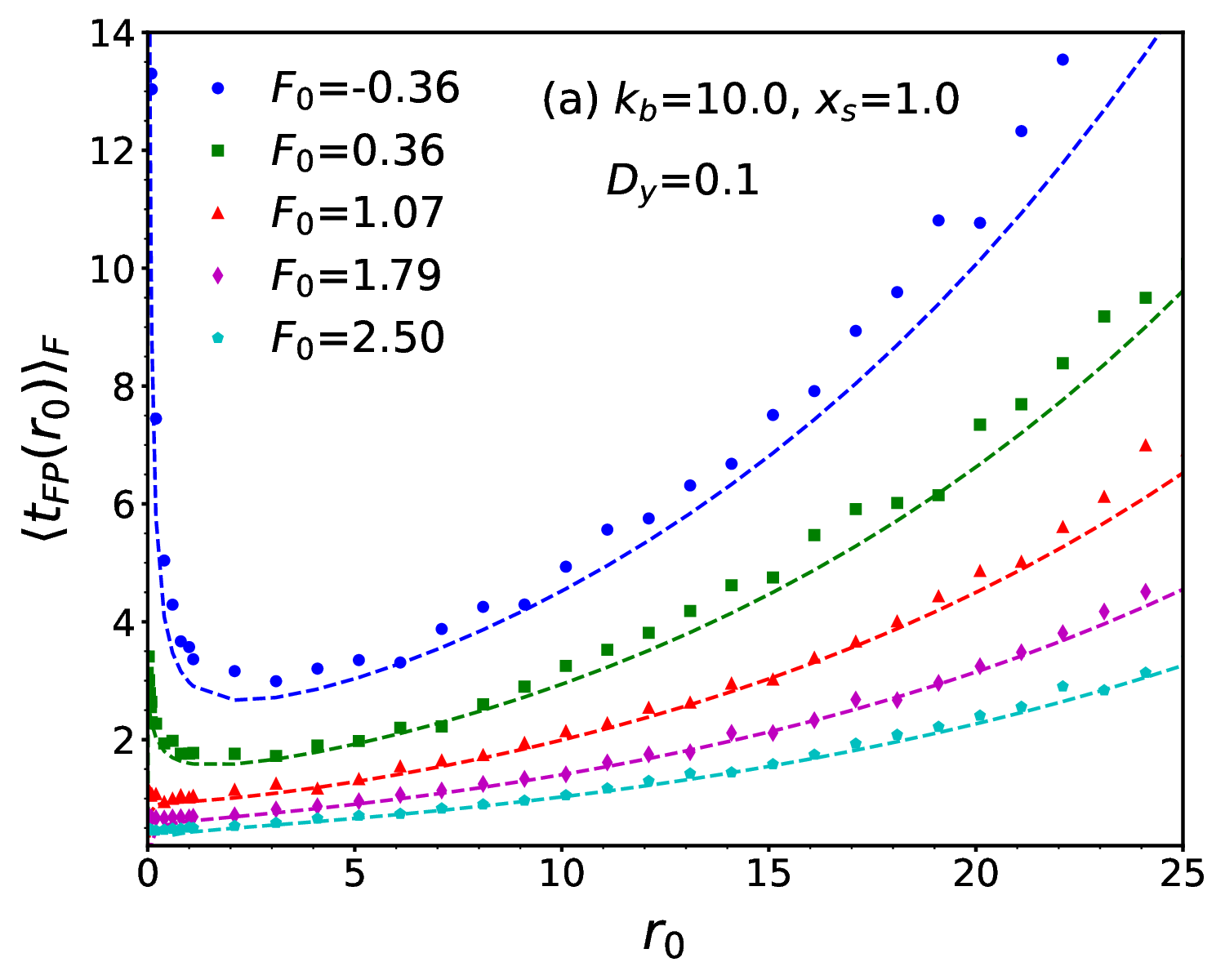}
    \end{subfigure}
    \hfill
    \begin{subfigure}[b]{0.495\textwidth}
        \centering
        \includegraphics[width=\textwidth]{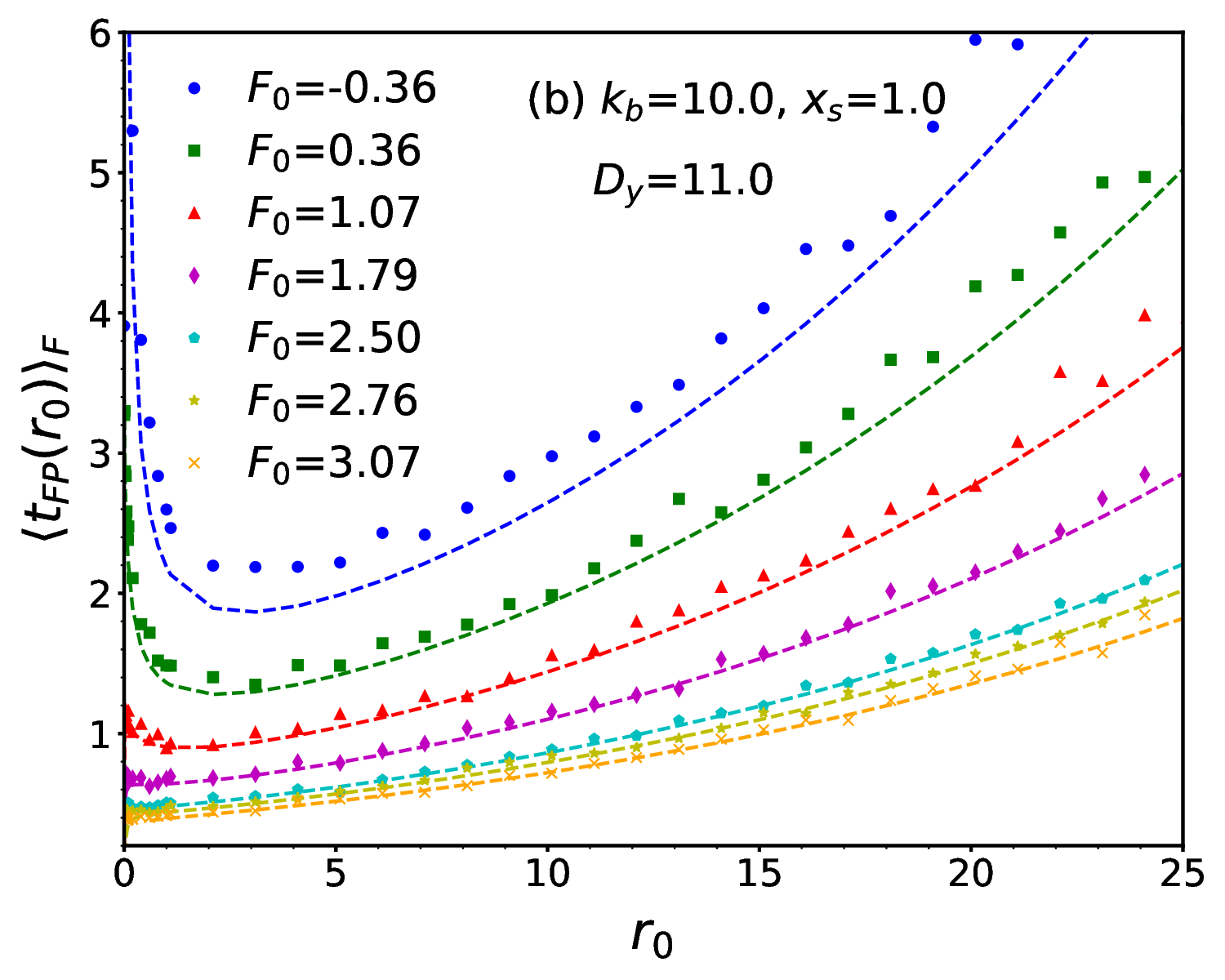}
    \end{subfigure}
        \vspace{0.2cm}
        \centering
    \begin{subfigure}[b]{0.495\textwidth}
        \centering
        \includegraphics[width=\textwidth]{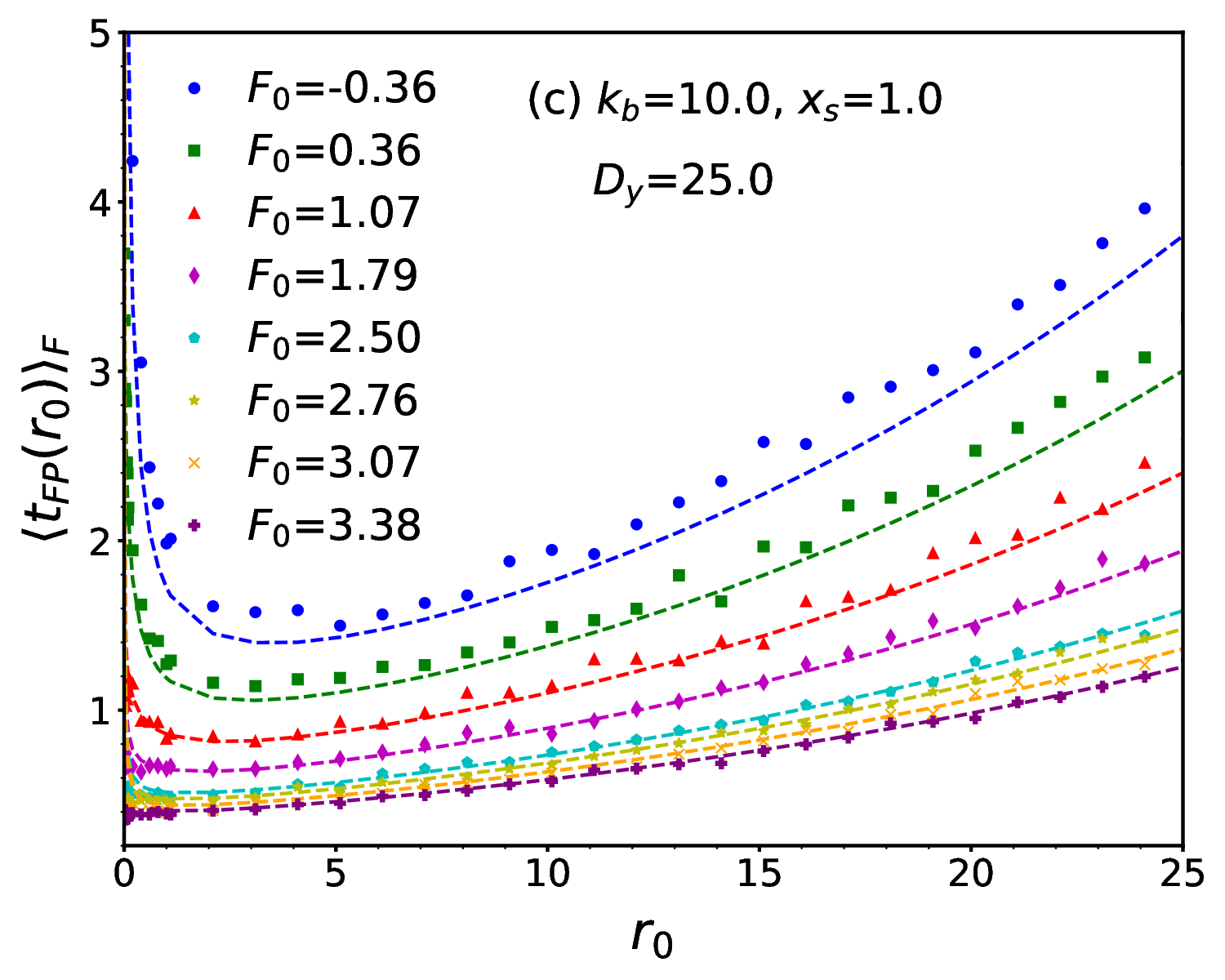}
    \end{subfigure}
\caption{Plot of the MFPT  under resetting and subject to external forces $F_0$ as a function of $r_0$ for three different values of $D_y$ across three panels corresponding to the three regimes: (a) $D_y < D_y^c$, (b) $D_y = D_y^c = 11.0$, and (c) $D_y > D_y^c$. The target (sink) is located at $x_s=1.0$,  with the tracer initially positioned at  $x(0)=0.$ The symbols represent simulation data, while the lines depict the analytical results derived from Eq. (\ref{tfp_force_0}). Here, $k_b=10.0,$ and other parameters are as specified in Appendix \ref{appen5}.} \label{fig:meantime_f0}
\end{figure*}

Suppose the tracer diffusing in the nonequilibrium bath is subjected to an external force  $+F_0$ along the positive $x$-direction, in addition to the previously discussed resetting mechanism. In the presence of an external linear potential $-F_0 x,$ a constant force $F_0$ is exerted along the positive $x$-direction. So, the dynamics of the tracer can be described by the following discrete stochastic equations, 
\begin{subequations}
\begin{align}
x(t+dt)=& x(0),\;\text{with probability } r_0 dt,\label{langevin-tracer_f0_reset1}\\
x(t+dt)=&x(t)-k_b \Big(x(t)-y(t)\Big)dt+\eta_x(t) dt + F_0 dt,\nonumber\\
&\qquad\qquad\;\text{with probability } (1-r_0 dt).\label{langevin-tracer_f0_reset2}
\end{align}
\end{subequations}
The target is located at $x=x_s>0$, and, the MFPT can be expressed as [see Eq. (\ref{tfp_force})]
\begin{align}
 \langle t_{FP}(r_0)\rangle_F=\frac{1}{r_0}\Bigl(\frac{ \widetilde{P}_F(x(0)=0,r_0|x(0)=0,0)}{ \widetilde{P}_F(x_s,r_0|x(0)=0,0)}-1\Bigr),\label{tfp_force_0}   
\end{align}
where the reset-free propagator in the presence of force $F_0$ is given by [see Eq. (\ref{pdf_px4_f0})]
\begin{align}
 P_F(x_f,t|0,0)=\sqrt{\frac{1}{4\pi\mathcal{C}_0^2(t)}} \exp\left(-\frac{\left(x_f- F_0\mathcal{S}_F(t)\right)^2}{4 \mathcal{C}_0^2(t)}\right). \label{pdf_px4_f00}
\end{align}
Here,  
\begin{subequations}
\begin{align}
&\mathcal{C}_0^2(t)=a_1+a_2 t + a_{11} e^{-k_b(1+\gamma)t} +a_{12} e^{-2k_b(1+\gamma)t},\label{C_0^2}\\
&F_0\mathcal{S}_F(t)=  
a_3+  a_4 t - a_3 e^{-k_b (\gamma + 1) t},\\
& a_3=\frac{F_0}{k_b (\gamma + 1)^2},\\
& a_4= F_0\frac{\gamma }{\gamma + 1}.
\end{align}
\end{subequations}
In the limit $k_b \gg 1,$ we approximate Eq (\ref{tfp_force_0}) to obtain [see Eq. (\ref{tfp_force_approx})] \begin{align}
\langle t_{FP}(r_0)\rangle_F \approx \frac{1}{r_0}\left[e^{-x_s\frac{a_4}{2a_2}+x_s\sqrt{\left(\frac{a_4}{2a_2}\right)^2+\frac{r_0}{a_2}}}-1\right]. \label{tfp_force_approx0}
\end{align} 
As in previous cases, an optimal resetting rate $r_0^*$ can be identified for any value of $F_0$, as shown in Fig. \ref{fig:meantime_f0}. For negative values of $F_0$, the force drives the particle away from the target, necessitating more frequent resetting to optimize the first-passage time. In this scenario, the MFPT diverges both without resetting and with infinite resetting. As the magnitude of the opposing force, $|F_0|$,  is lowered, the MFPT  also decreases.

When the direction of the force $F_0$ is reversed, aligned towards the target, the MFPT is further reduced since the force now pulls the tracer  towards the target. In such a case, the tracer is more likely to reach the target even with minimal resetting. At a certain critical force, $F_0^*$, the optimal resetting rate occurs at $r_0^*=0$, indicating that the tracer reaches the target in the shortest mean time even when it is not being reset stochastically to its initial position. For values of $F_0$ above this critical force, resetting prolongs the search time, thereby effectively slowing down the search process. This behavior is clearly depicted in panel (a) of  Fig. \ref{fig:rstar_f0}.

Across a  critical force $F_0^*$, two distinct regimes can be identified: (a) $r_0^*>0$ for $F_0<F_0^*$, and (b) $r_0^*=0$ for $F_0\geq F_0^*$. The optimal resetting rate $r_0^*$ occurs at $r_0=0$ when the following condition is satisfied \cite{Reuveni2016}:  \begin{align}
\langle t_{FP}^2\rangle_0= 2\langle t_{FP}\rangle_0^2,  \label{relation_momnet}  
\end{align}  where the moments of the first-passage time without resetting can be computed using the density given by Eq. (\ref{tfp_force_fpt}) as follows: $\langle t_{FP}^2\rangle_0=\frac{\partial^2}{\partial s^2}  \widetilde{f}_0(s) \Big|_{s=0}$ and $\langle t_{FP}\rangle_0=-\frac{\partial}{\partial s}  \widetilde{f}_0(s)\Big|_{s=0}.$ The exact analytical expression of the MFPT can be obtained in the limit of large $k_b$ values, as given by Eq. (\ref{tfp_force_approx0}). In this limit, we have $\langle t_{FP}^2\rangle_0=\frac{x_s^2}{a_4^2}+\frac{2 x_s a_2}{a_4^3}$ and $\langle t_{FP}\rangle_0=\frac{x_s}{a_4},$ and Using these results, the critical force is found to be 
\begin{align}
F_0^*=\frac{2 a_2}{x_s}\frac{\gamma + 1}{\gamma }=\frac{2}{x_s}\frac{D_x \gamma^2 + D_y}{\gamma(\gamma + 1)}.\label{critical_force}    
\end{align}
In panel (a) of Fig. \ref{fig:rstar_f0}, $F_0^*$ is estimated to be in close agreement with the analytically predicted value obtained from Eq. (\ref{critical_force}). In the thermal bath, $F_0^*=\frac{2 D_x}{x_s},$ which matches our case given in Eq. (\ref{critical_force}) when $D_y=\gamma D_x=D_y^{nc}$ \cite{Ahmad2019}. This result is not surprising, as at or beyond $F_0^*$, the MFPT is minimized without resetting, and in the absence of resetting $(r_0=0)$, the condition $D_y=D_y^{nc}$ characterizes a thermal bath. When interacting strongly with hot particles ($D_y>D_y^{nc}$), a comparatively higher $F_0^*$ is required than when coupled with cold particles ($D_y<D_y^{nc}$).
Before reaching $F_0^*$, the MFPT at $r_0^*$ decreases with $F_0$ following an exponential decay $\exp\left(-F_0/\Bar{F}_0\right)$, as shown in panel (b) of Fig. \ref{fig:rstar_f0}. For lower $D_y$, this decay is faster, with a smaller $\Bar{F}_0$ value. Beyond $F_0^*$, all cases appear to follow a decay law with same $\Bar{F}_0$.

\begin{figure*}[htp]
    \begin{subfigure}[b]{0.495\textwidth}
        \caption[]{\raggedright \small }
        \includegraphics[width=\textwidth]{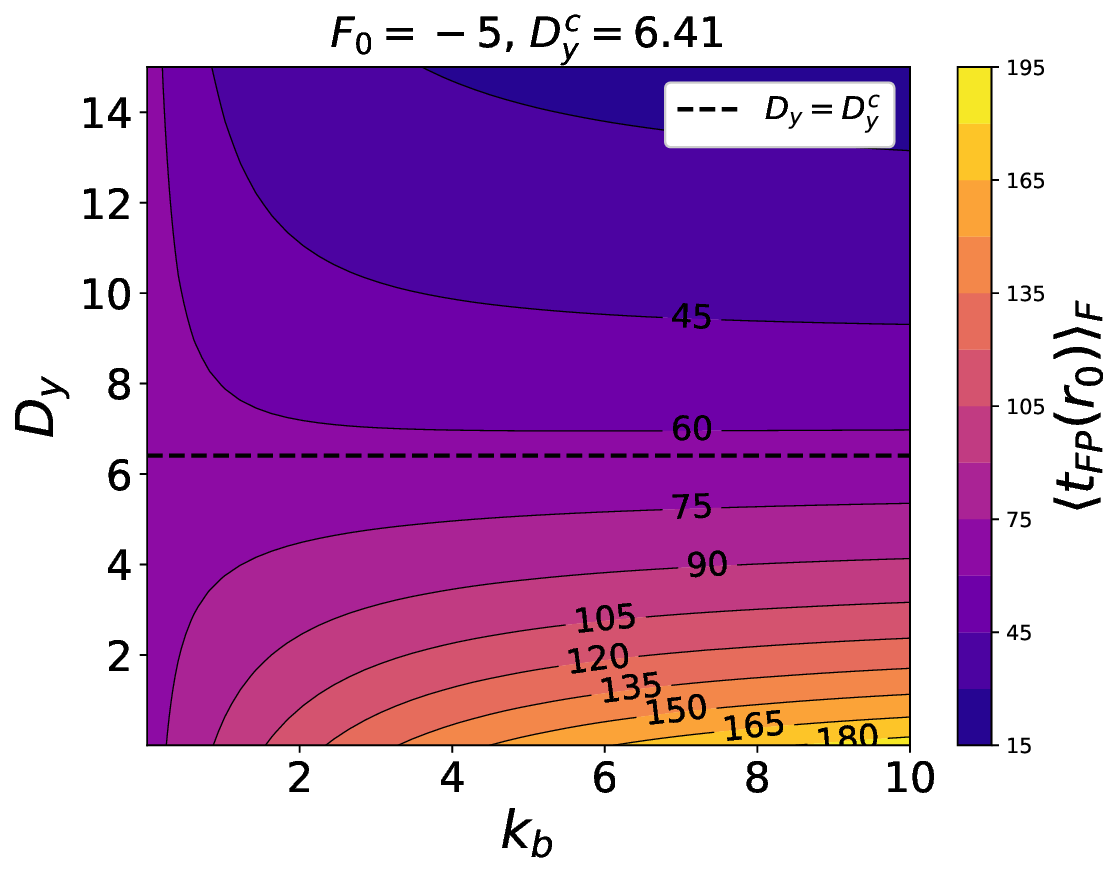}
    \end{subfigure}
    \hfill
    \begin{subfigure}[b]{0.495\textwidth}
        \caption[]{\raggedright \small }
        \includegraphics[width=\textwidth]{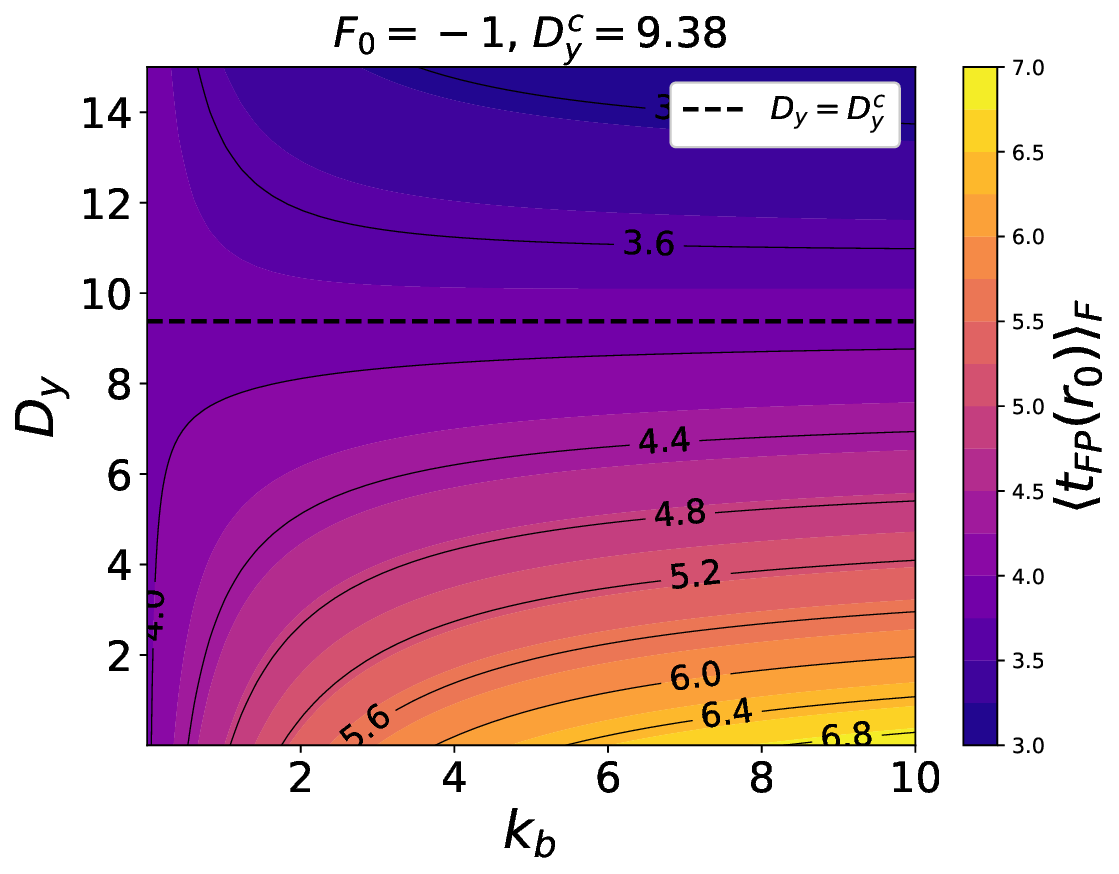}
    \end{subfigure}
    \hfill
    \begin{subfigure}[b]{0.495\textwidth}
        \caption[]{\raggedright \small }
        \includegraphics[width=\textwidth]{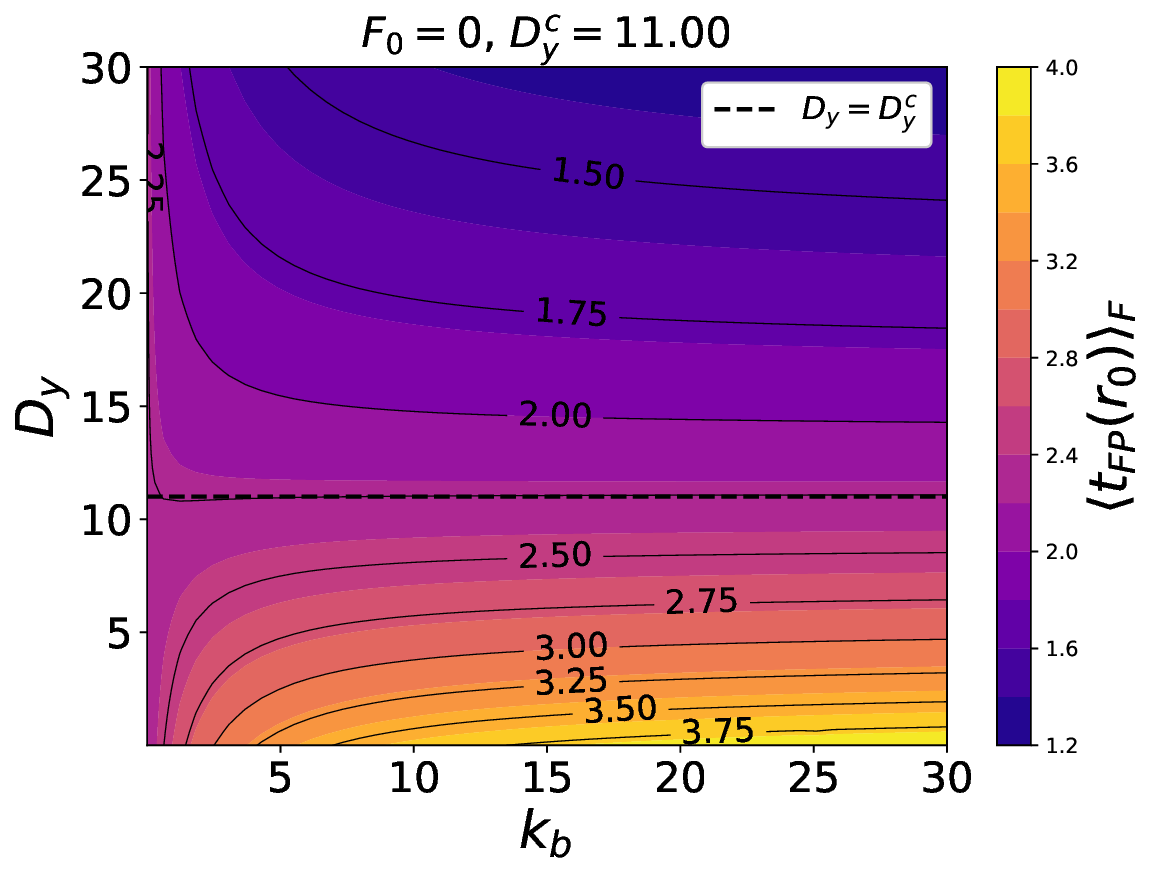}
    \end{subfigure}
    \vspace{0.2cm}
    \hfill
    \begin{subfigure}[b]{0.495\textwidth}
        \caption[]{\raggedright \small }
        \includegraphics[width=\textwidth]{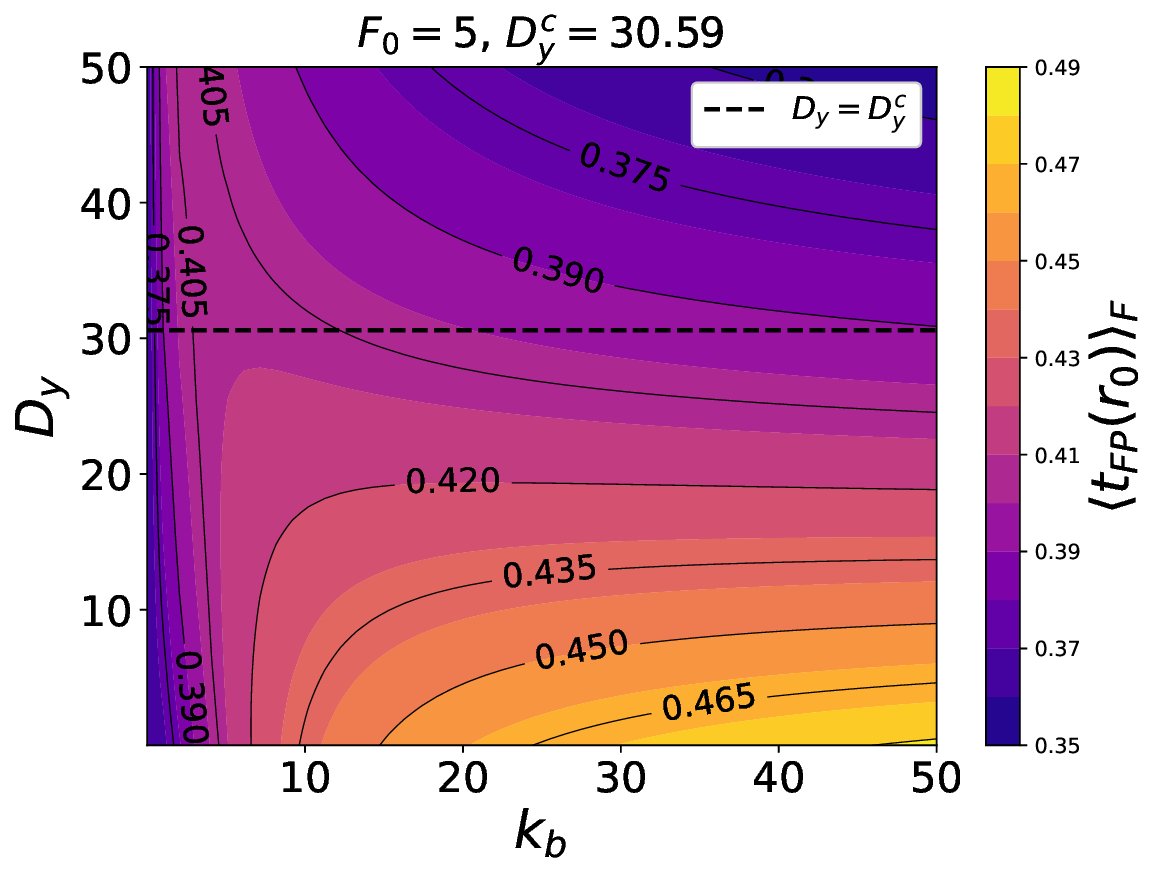}
    \end{subfigure}  
\caption{Contour plot of the MFPT computed from Eq. (\ref{tfp_force_0}) under resetting and subject to external forces $F_0$  shown as a function of  $D_y$ and  $k_b$ across four panels for different values of $F_0$. The target is located at $x_s=1.0$,  and the tracer is reset stochastically at a rate $r_0=10$ to its initial position  $x(0)=0.$ Additional details are as specified in Appendix \ref{appen5}.} \label{fig:contour_kb_dy_f0}
\end{figure*}

 At a finite, non-zero $r_0$, there exists a specific diffusivity, denoted as $D_y^c$ following previous conventions, for which the MFPT remains the same at both very low and very high coupling strengths $k_b$. However, $D_y^c$ now depends on both $r_0$ and $F_0$, and is given by [see Eq. (\ref{Dy_regime2})]:
\begin{align}
 D_y^c=D_x(2\gamma+1)+\frac{F_0}{2r_0}(\gamma+1)\left[F_0+\sqrt{F_0^2+4 D_x r_0}\right].\label{Dy_regime20}  
\end{align}
As shown in panel (d) of Fig. \ref{fig:contour_kb_dy_f0}, $D_y^c$ does not imply that the MFPT for $D_y=D_y^c$ is independent of the coupling strength $k_b$, unlike in the previous force-free case, particularly for positive values of $F_0$. In line with the previous case, regime II is defined when $D_y=D_y^c$, with regimes I and III emerging when $D_y<D_y^c$ and $D_y>D_y^c$, respectively. However, the characteristics of these regimes differ significantly from the force-free scenario.  
As illustrated in Fig. \ref{fig:contour_kb_dy_f0}, in regime I, where the bath particles are cold $(D_y<D_y^c)$, very strong interactions increase the MFPT compared to very weak coupling. In contrast, in regime III, where the bath particles are hot $(D_y>D_y^c)$, the opposite trend is observed. However, the variation of MFPT with respect to $k_b$ is not necessarily monotonic, especially when $F_0$ is directed toward the target. Specifically, as shown in panel (d), the region of maximum MFPT values shifts toward lower $k_b$ values, with the peak becoming more pronounced as the system transitions from regime I to regime III.

From Eq. (\ref{Dy_regime20}), we recover the force-free case ($F_0=0$) with $D_y^c=D_x(2\gamma+1)$. In the extreme case of a large negative force (directed away from the target),  $D_y^c=D_y^{nc}$, representing the thermal limit in a reset-free condition, which sets a lower bound on the diffusivity defining cold particles. For a force directed toward the target, $D_y^c$ scales with $F_0^2$, indicating that as the force increases, stronger coupling with bath particles significantly hinders the search process.

\section{Conclusions  \label{conclusion}}
In this work, we have explored the influence of stochastic resetting on the dynamics of a tracer particle in a nonequilibrium bath characterized by harmonic coupling with bath particles. At a particular diffusivity of bath particle, denoted as $D_y^c$, the coupling strength $k_b$ has a minimal effect on the tracer dynamics. Diffusivities below and above $D_y^c$ characterize  cold and hot  baths, respectively. Our analysis reveals that, under resetting, the tracer exhibits slower relaxation during transient times when strongly coupled to a cold bath, whereas it achieves a steady state more rapidly when strongly interacting with a hot bath. This behavior can be attributed to the fact that stronger coupling to hot particles results in larger displacements, a behavior relevant in activity-induced environments \cite{Netz2018, Goswami2023}. In contrast, coupling to cold particles confines the tracer more, a characteristic feature of viscoelastic media \cite{Guevara-Valadez2023}. In the extreme cases of very strong or very weak coupling, the  steady-state distribution of the tracer is represented by a Laplace distribution with an effective diffusivity $a_2$. For very weak coupling, $a_2$ is replaced by the tracer's bare diffusivity $D_x$, corresponding to the thermal case. In contrast, at very strong coupling $a_2=D_x$ only when $D_y=D_y^c$.

In line with the dynamical behavior, a similar pattern is reflected in the presence of an immobile target:  at a given finite resetting rate, stronger coupling with hot particles minimizes the MFPT, whereas the opposite trend is observed in a cold bath. In the presence of an external linear potential, we identified a critical force $F_0^*$ beyond which resetting is no longer necessary for the optimal search strategy, signifying a resetting transition \cite{Biswas2023}. 
Theoretically, we find that higher $F_0^*$ is required for larger $D_y$. In the force-driven case, the characteristics of both cold and hot baths differ from the force-free scenario: The MFPT exhibits a nonlinear dependence on the coupling strength, especially when the force is directed towards the target. In such cases, stronger coupling impedes the search process. In contrast, when the force opposes the target direction, stronger coupling becomes essential for facilitating the search.

Our findings present opportunities for experimental validation in viscoelastic and active environments and can be extended to more complex systems exhibiting non-Markovian dynamics with anomalous diffusion \cite{Metzler2000, Godec2014, Kimura2023,harbola2024}. These systems often incorporate memory effects with power-law characteristics \cite{Metzler2000, Kimura2023}, which could be integrated into our framework using the Markov embedding method \cite{Goychuk2009}. Additionally, incorporating active tracer dynamics within our formalism would expand its applicability to a wider array of systems influenced by active fluctuations \cite{Guevara-Valadez2023, Das2023,goswami2019heat,Abbasi2023}. Future work could also explore the thermodynamic aspects of our model, providing further insights into energy exchanges and entropy production in these non-equilibrium settings \cite{Lahiri2024,Kumari2024}.



\section*{Acknowledgments}
This work was supported by the National Science and Technology Council, the Ministry of Education (Higher Education Sprout Project NTU-113L104022), and the National Center for Theoretical Sciences of Taiwan.

\appendix
\setcounter{figure}{0}
\begin{widetext}
\section{Derivation of the PDF \label{appen1}}
Here, we present an analytical approach to derive the PDF for the process discussed in Sec. \ref{sec2}. 
\subsection{Free particle}
By taking the Laplace transform of Eqs. \ref{langevin-tracer}-\ref{langevin-bath}, we arrive at
\begin{subequations}
\begin{align}
s\,\widetilde{x}(s)-x(0)=&-\sum_{i=1}^{N}\,k_b \Big(\widetilde{x}(s)-\widetilde{y}_i(s)\Big)+\widetilde{\eta}_x(s),\label{langevin-tracer_lap}\\
s\,\widetilde{y}_i(s)-y_i(0)=&-\gamma\,k_b \Big(\widetilde{y}_i(s)-\widetilde{x}(s)\Big)+\widetilde{\eta}_y(s).\label{langevin-bath_lap}
\end{align}
\end{subequations}
Rearranging Eq. (\ref{langevin-bath_lap}), we can express $y_i(s)$ as 
$\widetilde{y}_i(s)=\frac{y_i(0)}{s+\gamma\,k_b}+\gamma\,k_b\frac{\widetilde{x}(s)}{s+\gamma\,k_b}+\frac{\widetilde{\eta}_y(s)}{s+\gamma\,k_b}.$
Using this expression in Eq. (\ref{langevin-tracer_lap}), we obtain
$$\widetilde{x}(s)\left[s+N k_b-\frac{N\gamma\,k_b^2}{s+\gamma\,k_b}\right]=x(0)+\sum_{i=1}^{N}\,k_b\frac{y_i(0)}{s+\gamma\,k_b}+N k_b\,\frac{\widetilde{\eta}_y(s)}{s+\gamma\,k_b}+\widetilde{\eta}_x(s).$$
Rearranging the terms, we can recast this as
$$\widetilde{x}(s)=x(0)\frac{s+\gamma\,k_b}{s\left[s+k_b(N+\gamma)\right]}+\sum_{i=1}^{N}\,k_b\frac{y_i(0)}{s\left[s+k_b(N+\gamma)\right]}+\sum_{i=1}^{N}\, k_b\,\frac{\widetilde{\eta}_y(s)}{s\left[s+k_b(N+\gamma)\right]}+\widetilde{\eta}_x(s)\frac{s+\gamma\,k_b}{s\left[s+k_b(N+\gamma)\right]}.$$
Taking the inverse Laplace transform, the above equation takes the form
\begin{align}
 x(t)= x(0)\Psi(t)+\sum_{i=1}^{N}\Phi(t) y_i(0)+\int_{0}^{t}dt_1\,\Psi(t-t_1)\eta_x(t_1)
 + \sum_{i=1}^{N}\,\int_{0}^{t}dt_1\,\Phi(t-t_1)\eta_y(t_1),
\end{align}
where we define the two terms as
\begin{subequations}
  \begin{align}
    & \Psi(t)=\frac{\left[\gamma+N\,e^{-k_b(N+\gamma)t}\right]}{N+\gamma}, \label{psi}\\
    & \Phi(t)=\frac{\left[1-e^{-k_b(N+\gamma)t}\right]}{N+\gamma}. \label{phi}
  \end{align}  
\end{subequations}
Using the above result, the PDF of $x$ can be expressed as \cite{goswami2019heat,goswami2023j}
\begin{align}
& P_0(x,t|x(0),0)=\langle \delta\left(x-x(t)\right)\rangle =\frac{1}{2\pi}\int dq_x\,e^{-i q_x x}\langle e^{i q_x x(t) }\rangle_{x(t)}\nonumber\\
&=\frac{1}{2\pi}\int dq_x\,e^{-i q_x x}\langle e^{i q_x x(0) \Psi(t) }\rangle_{x(0)}
\prod_{i=1}^{N}\langle e^{i q_x \Phi(t) y_{i}(0) }\rangle_{y_i(0),x(0)} \prod_{i=1}^{N}\langle e^{i q_x \int_{0}^{t}dt_1\,\Phi(t-t_1)\eta_{y}(t_1)}\rangle_{\eta_{y}}\,\langle e^{i q_x  \int_{0}^{t}dt_1\,\Psi(t-t_1)\eta_x(t_1)}\rangle_{\eta_x}\label{pdf_px},
\end{align}
where $\langle\cdots\rangle_{z(t)}$ represents the ensemble average over
all realisations of variable $z(t)$.  
For the delta-correlated Gaussian noise $\eta(t)$ with its variance $D_T$, its characteristic functional is given by \cite{van1992stochastic,goswami2019diffusion} 
\begin{align}
\left\langle\exp\left[i\int_{0}^{t}\,dt'p(t') \eta(t')\right]\right\rangle_{\eta}=\exp\left(- D_T\int_{0}^{t}dt_1\,p^2(t_1)\right).\label{charac_Gaussian}
\end{align}
Using this relation in Eq. (\ref{pdf_px}), the PDF can be computed as follows
\begin{align}
&P_0(x,t|x(0),0) \nonumber\\
&= \frac{1}{2\pi}\int dq_x\,e^{-i q_x \left(x- x(0) \Psi(t)\right)}\,e^{-D_x q_x^2\int_{0}^{t}dt_1\,\Psi^2(t-t_1)-N D_y q_x^2\int_{0}^{t}dt_1\,\Phi^2(t-t_1)}\prod_{i=1}^{N}\int dy_i(0) P(y_i(0))\,e^{i q_x \Phi(t) y_{i}(0) }\nonumber\\
  &=\frac{1}{2\pi}\sqrt{\frac{\gamma k_b}{2\pi D_y}} \prod_{i=1}^{N}\int dy_i(0)\int dq_x\,e^{-i q_x \left(x- x(0) \Psi(t)\right)+i q_x \Phi(t) y_{i}(0)}\,e^{-D_x q_x^2\int_{0}^{t}dt_1\,\Psi^2(t-t_1)-N D_y q_x^2\int_{0}^{t}dt_1\,\Phi^2(t-t_1)-\frac{\gamma k_b}{2D_y}y_i^2(0)}\nonumber\\
  &= \frac{1}{2\pi}\int dq_x\,e^{-i q_x \left(x- x(0) \Psi(t)\right)-q_x^2\xi_0^2(t)}\nonumber\\
  &=\sqrt{\frac{1}{4\pi\xi_0^2(t)}} \exp\left(-\frac{\left(x- x(0) \Psi(t)\right)^2}{4\xi_0^2(t)}\right),\label{pdf_px4}
\end{align}
where 
\begin{align}
  \xi_0^2(t)=& \frac{N D_y \left[1+e^{-2k_b(N+\gamma)t}-2 e^{-k_b(N+\gamma)t}\right]}{2\gamma k_b(N+\gamma)^2}+\frac{N D_y}{(N+\gamma)^2}\left[t-\frac{3}{2k_b(N+\gamma)}+\frac{2\,e^{-k_b(N+\gamma)t}}{k_b(N+\gamma)}-\frac{e^{-2k_b(N+\gamma)t}}{2k_b(N+\gamma)}\right]\nonumber\\
  &+\frac{D_x}{(N+\gamma)^2} \left[\gamma^2 t+\frac{N^2}{2k_b(N+\gamma)}+\frac{2\gamma N}{k_b(N+\gamma)}-\frac{2\gamma N\,e^{-k_b(N+\gamma)t}}{k_b(N+\gamma)}-\frac{N^2 e^{-2k_b(N+\gamma)t}}{2k_b(N+\gamma)}\right]\nonumber\\
  &=\alpha_1+\alpha_2 t + \alpha_{11} e^{-k_b(N+\gamma)t} +\alpha_{12} e^{-2k_b(N+\gamma)t}.\label{xi_0^2}
\end{align}
Let us define the following parameters
\begin{subequations}
\begin{align}
& \alpha_1=\frac{N \left[D_y (N-2 \gamma) + D_x \gamma (N+4\gamma)\right]}{2 \gamma k_b (\gamma + N)^3}, \label{alpha1}\\
& \alpha_2=\frac{D_x \gamma^2 + D_y N}{(\gamma + N)^2},\label{alpha2}\\
& \alpha_{11}=-\frac{N \left[2D_y (N-\gamma)+4 D_x \gamma^2 \right]}{2 \gamma k_b (\gamma + N)^3},\label{alpha11}\\
& \alpha_{12}=\frac{N^2 \left[D_y -D_x \gamma \right]}{2 \gamma k_b (\gamma + N)^3}\label{alpha12}.
\end{align}    
\end{subequations}
Using $P_0(x,t|x(0),0)$  [Eq. (\ref{pdf_px4})] in Eq. (\ref{last_renewal_equation}), one can determine the PDF of $x$ in the presence of resetting. Note that the PDF in Eq. (\ref{last_renewal_equation}) cannot be calculated exactly by analytical means. Therefore, we computed the PDF numerically, as shown in Fig. \ref{fig:pdf_reset}. In the large-time limit, the PDF can be expressed as
\begin{align}
P_{r,st}(x)=\lim_{t \rightarrow \infty}P_r(x,t|0,0) = r_0 \int_{0}^{t}d\tau e^{-r_0 \tau}\,P_0(x,\tau), \label{pdf_reset_st} 
\end{align}
which is essentially the Laplace transform of the reset-free propagator with respect to the variable $r_0$, multiplied by $r_0.$ Given that the transformation is not analytically tractable, we consider the large $k_b$ limit, for which the exponential terms occurring in $\xi_0^2(t)$ can be ignored, i.e., $e^{- k_b (\gamma + N) t - r_0 t} \rightarrow 0,\,e^{-2 k_b (\gamma + N) t - r_0 t} \rightarrow 0.$ This leads to $\xi_0^2(t) \approx \alpha_1+\alpha_2 t.$  Thus, for $\alpha_1>0,$ the PDF approximates to
\begin{align}
 P_{r,st}(x) \approx \frac{1}{4} \sqrt{\frac{r_0}{\alpha_2}}\left[
\exp\left(-x \sqrt{\frac{r_0}{\alpha_2}} + \frac{r_0 \alpha_1}{\alpha_2}\right) 
\operatorname{Erfc}\left(-\frac{x}{2 \sqrt{\alpha_1}} + \sqrt{\frac{r_0 \alpha_1}{\alpha_2}}\right)+ 
\exp\left(x \sqrt{\frac{r_0}{\alpha_2}} + \frac{r_0 \alpha_1}{\alpha_2}\right) 
\operatorname{Erfc}\left(\frac{x}{2 \sqrt{\alpha_1}} + \sqrt{\frac{r_0 \alpha_1}{\alpha_2}}\right)\right].\label{pdf_reset_st_approx} 
\end{align}
In the limit $k_b \rightarrow \infty,$the PDF simplifies to 
\begin{align}
 P_{r,st}(x) \approx \frac{1}{2} \sqrt{\frac{r_0}{\alpha_2}} e^{-|x| \sqrt{\frac{r_0}{\alpha_2}}}, \label{pdf_reset_st_approx_kbl}  
\end{align} which is similar to the thermal case \cite{Evans2020}, with the difference being that the thermal diffusivity $D_x$ is replaced by $\alpha_2=\frac{D_x \gamma^2 + D_y N}{(\gamma + N)^2}.$ 
Note that Eq. (\ref{pdf_reset_st_approx_kbl}) is valid for any value  of $D_y$ in the limit $k_b \gg 1.$ When   $D_y=D_y^c=D_x(2\gamma+N)$, we obtain $\alpha_2=D_x$, otherwise $\alpha_2 \neq D_x$. For $D_y<D_y^c$, $\alpha_2<D_x$, and for  $D_y>D_y^c$,  $\alpha_2>D_x.$ 

For $k_b\rightarrow 0$,  one recovers the thermal case, where  $\xi_0^2(t) \approx D_x t$, and therefore, the PDF simplifies to  \begin{align}
 P_{r,st}(x) \approx \frac{1}{2} \sqrt{\frac{r_0}{D_x}} e^{-|x| \sqrt{\frac{r_0}{D_x}}} \label{pdf_reset_st_approx_thermal}.  
\end{align}

The MSD can be computed from the propagator using the relation
\begin{align}
  \langle x^2(t)\rangle_0=-\frac{\partial^2}{\partial q_x^2} \hat{P}_0(q_x,t|x(0)=0,0)\big|_{q_x=0},\;\langle x^2(t)\rangle_r=-\frac{\partial^2}{\partial q_x^2} \hat{P}_r(q_x,t|x(0)=0,0)\big|_{q_x=0}, \label{msd_fourier}
\end{align}
where $\hat{P}(q_x,t|x(0)=0,0)$ denotes the Fourier transform of  $P(x,t|x(0)=0,0).$ From Eq. (\ref{pdf_px4}), the Fourier transform of the reset-free propagator can be written as
\begin{align}
 \hat{P}_0(q_x,t|x(0)=0,0)=  \mathcal{F}\left( P_0(x,t|x(0)=0,0)\right)= \exp\left(-q_x^2\xi_0^2(t)\right).
\end{align}
Using Eq. (\ref{msd}) and Eq. (\ref{xi_0^2}), one can write the MSD as 
\begin{align}
\langle x^2(t)\rangle_r = 2 e^{-r_0 t} \xi_0^2(t) + 2r_0  \xi_{r0}^2(t),\label{msd_exact}
\end{align}
where
\begin{align}
\xi_{r0}^2(t)&=\int_{0}^{t}d\tau e^{-r_0 \tau}\, \xi_0^2(\tau)\nonumber\\
&=\frac{\alpha_1}{r_0}(1-e^{-r_0 t})+\frac{\alpha_2}{r_0^2}[1-(1+r_0 t)e^{-r_0 t}] +\frac{\alpha_{11}}{r_0+k_b(N+\gamma)}(1-e^{-[r_0+k_b(N+\gamma)] t})\nonumber\\
& +\frac{\alpha_{12}}{r_0+2 k_b(N+\gamma)}(1-e^{-[r_0+2 k_b(N+\gamma)] t}).\label{xir0}    
\end{align}
In the short-time limit, i.e., for $t\rightarrow 0$, the MSD approximates to
\begin{align}
  \langle x^2(t)\rangle_r \approx  2D_x t.\label{msd_shortt}
\end{align}
In the long-time limit, the MSD converges to a fixed value given by
\begin{equation}
 \langle x^2(t)\rangle_r \approx 2 \alpha_1+\frac{2 \alpha_2}{r_0}+\frac{2r_0\alpha_{11}}{r_0+k_b(N+\gamma)}+\frac{2 r_0\alpha_{12}}{r_0+2 k_b(N+\gamma)}.\label{msd_longt}
\end{equation}

\subsection{Particle subjected to a time-independent force $F_0$}
In the presence of external force $F_0$, from Eq. (\ref{langevin-tracer_f0_reset2}), one can write 
\begin{align}
 x(t)= x(0)\Psi(t)+\Phi(t) y_i(0)+\int_{0}^{t}dt_1\,\Psi(t-t_1)\eta_x(t_1)+ \int_{0}^{t}dt_1\,\Phi(t-t_1)\eta_y(t_1)+F_0\int_{0}^{t}dt_1\,\Psi(t-t_1).
\end{align}
Similar to Eq. (\ref{pdf_px4}), the reset-free propagator in the presence of external force $F_0$ can be expressed as 
\begin{align}
P_{F}(x,t|x(0)=0,0) &=   \frac{1}{2\pi}\int dq_x\,e^{-i q_x x+i q_x F_0 \Psi_F(t)}\,\exp\left(-q_x^2\xi_0^2(t)\right)\nonumber\\
  &=\sqrt{\frac{1}{4\pi\xi_0^2(t)}} \exp\left(-\frac{\left(x-F_0 \Psi_F(t)\right)^2}{4\xi_0^2(t)}\right),\label{pdf_px4_f0}
\end{align}
where 
\begin{align}
\Psi_F(t)=\int_{0}^{t} dt_1 \Psi(t_1)=\frac{N}{k_b (\gamma + N)^2} 
+ \frac{\gamma }{\gamma + N}t - \frac{N e^{-k_b (\gamma + N) t}}{k_b (\gamma + N)^2}. 
\end{align}

In the limit $k_b(N+\gamma)\gg 1,$ one can approximate $F_0\Psi_F(t) \approx \alpha_3 +\alpha_4 t,$ where
\begin{subequations}
\begin{align}
& \alpha_3=F_0\frac{N}{k_b (\gamma + N)^2},  \label{alpha3}\\
& \alpha_4=F_0\frac{\gamma }{\gamma + N}.\label{alpha4}
\end{align}    
\end{subequations}

The propagator under resetting cannot be computed exactly using the renewal equation [\ref{last_renewal_equation}]. In the long-time limit, the propagator can be found by taking the Laplace transform of $P_{F}(x,t|x(0),0)$. In the limit $k_b(N+\gamma)\gg 1,$ one can compute the Laplace transform of the propagator $P_{F}(x,t|x(0),0)$ using Eq. (\ref{p_f}) by replacing $c_i$ with $\alpha_i,$ where $i \in (1,4)$, i.e., 
\begin{align}
\widetilde{P}_F(x,r_0|x(0)=0,0)=\int_{-\infty}^{+\infty}\,\frac{dq}{2\pi}\,e^{-i q x}\,\frac{e^{-\alpha_1 q^2+i \alpha_3 q}}{\alpha_2 q^2-i \alpha_4 q+ r_0}.   \label{pdf_px4_f0_laplace}   
\end{align}
In the limit $k_b \gg 1,$ one can assume $\alpha_1 \rightarrow 0$ and $\alpha_3 \rightarrow 0.$ So, the PDF without resetting in the presence of force $F_0$ can be approximated as  
\begin{align}
\widetilde{P}_F(x_s,r_0|x(0)=0,0) \approx \int_{-\infty}^{+\infty}\,\frac{dq}{2\pi}\,\frac{e^{-i q x_s}}{\alpha_2 q^2-i \alpha_4 q+ r_0} \approx \frac{e^{x_s\frac{\alpha_4}{2\alpha_2}-x_s\sqrt{\left(\frac{\alpha_4}{2\alpha_2}\right)^2+\frac{r_0}{\alpha_2}}}}{\sqrt{\alpha_4^2 + 4 \alpha_2 r_0}} . \label{pf_noreset}
\end{align}
In the limit $k_b\ll 1,$ we can make the approximations  $\xi_0^2(t) \approx D_x t$ and $\Psi_F(t) \approx t.$ Thus, the Laplace transform of the propagator simplifies to
\begin{align}
 \widetilde{P}_F(x_s,r_0|x(0)=0,0) \approx \int_{-\infty}^{+\infty}\,\frac{dq}{2\pi}\,\frac{e^{-i q x_s}}{D_x q^2-i F_0 q+ r_0} \approx \frac{e^{x_s\frac{F_0}{2 D_x}-x_s\sqrt{\left(\frac{F_0}{2D_x}\right)^2+\frac{r_0}{D_x}}}}{\sqrt{F_0^2 + 4 D_x r_0}} . \label{pf_noreset_0}  
\end{align}

In the presence of resetting and force, the MSD in the long-time limit can be expressed as [see Eq. (\ref{msd})]
\begin{align}
\langle x^2(t)\rangle_{F,r} \approx  r_0 \int_{0}^{t}d\tau e^{-r_0 \tau}\,\langle x^2(\tau)\rangle_F,
\end{align}
where, by virtue of Eq. (\ref{msd_fourier}) and Eq. (\ref{pdf_px4_f0}), we have 
\begin{align}
  \langle x^2(t)\rangle_F=-\frac{\partial^2}{\partial q_x^2} \hat{P}_F(q_x,t|x(0)=0,0)\big|_{q_x=0}=2\xi_{0}^2(t) + F_0^2 \Psi_F^2(t).
\end{align}
Thus, the MSD in the limit $t \rightarrow \infty$ can be computed as
\begin{align}
 \langle x^2(t)\rangle_{F,r} & = 2 \alpha_1+\frac{2 \alpha_2}{r_0}+\frac{2r_0\alpha_{11}}{r_0+k_b(N+\gamma)}+\frac{2 r_0\alpha_{12}}{r_0+2 k_b(N+\gamma)}  + \frac{2 \alpha_4^2}{r_0^2}+\frac{2\alpha_3 \alpha_4}{r_0}+ \alpha_3^2\nonumber\\
 & + \frac{2r_0 \alpha_3 \alpha_4}{\left(k_b (\gamma + N)+r_0\right)^2}+\frac{2 r_0 \alpha_3^2}{k_b (\gamma + N)+r_0}+\frac{r_0\alpha_3^2}{2k_b (\gamma + N)+r_0}.\label{x2fr}
\end{align}
The above can be approximated to  
\begin{align}
\langle x^2(t)\rangle_{F,r} &
\approx \begin{cases}
& \frac{2}{r_0}\left[D_x+\frac{F_0^2}{r_0}\right] ,\; k_b \rightarrow 0\nonumber\\
& \frac{2}{r_0}\left[\frac{D_x \gamma^2 + D_y N + \frac{F_0^2 \gamma^2}{r_0}}{(\gamma + N)^2}\right],\; k_b \rightarrow \infty
\end{cases}\\ \label{msd_f0_st},
\end{align}
indicating that the steady-state MSD, $\langle x^2(t)\rangle_{F,r}$, depends on the magnitude of the external force $F_0$ and remains independent of its direction.
\end{widetext}

\section{Relaxation to steady state \label{appen2}}

 \begin{figure*}[htp]
    \centering
    \begin{subfigure}[b]{0.495\textwidth}
        \centering
        \includegraphics[width=\textwidth]{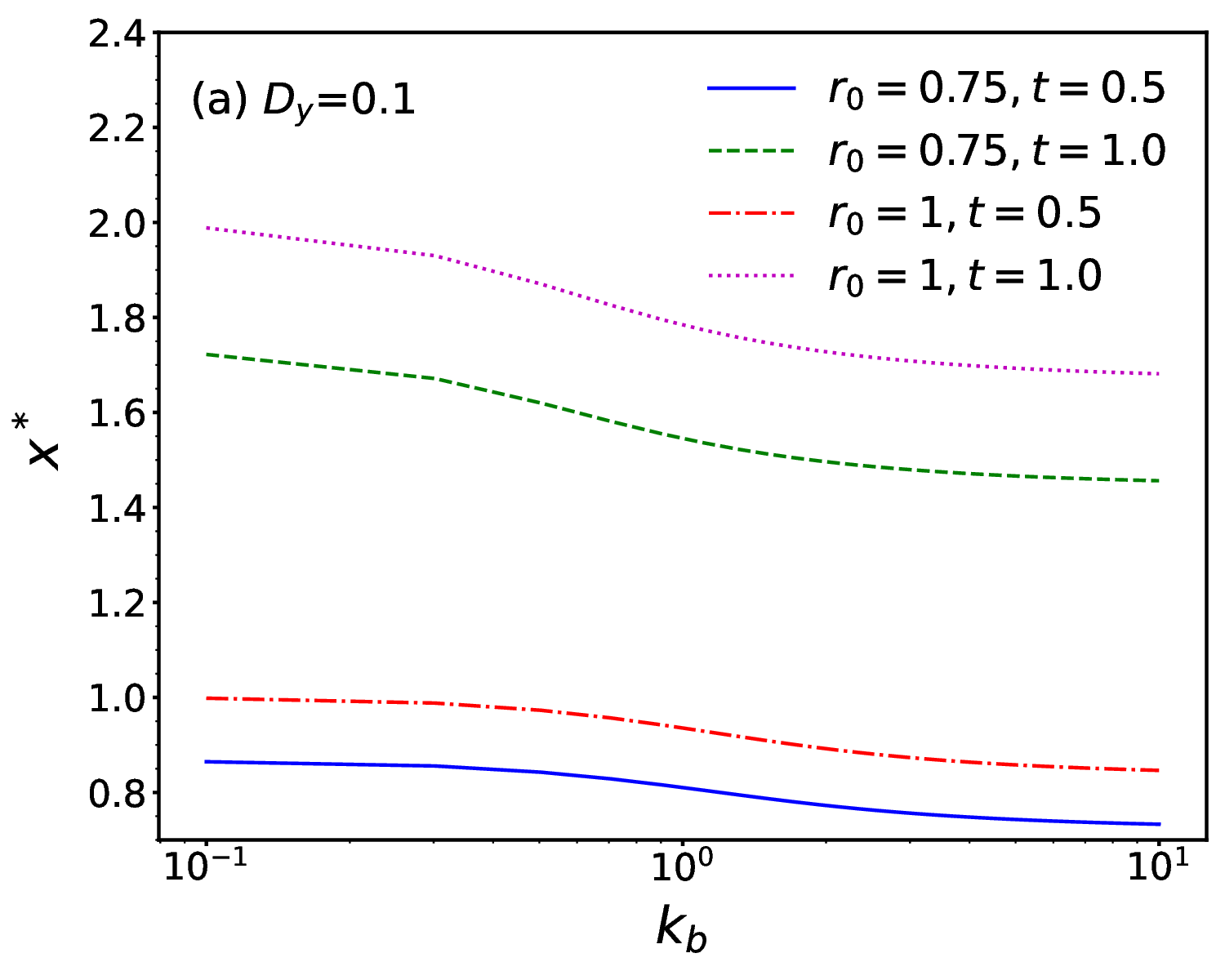}
    \end{subfigure}
    \hfill
    \begin{subfigure}[b]{0.495\textwidth}
        \centering
        \includegraphics[width=\textwidth]{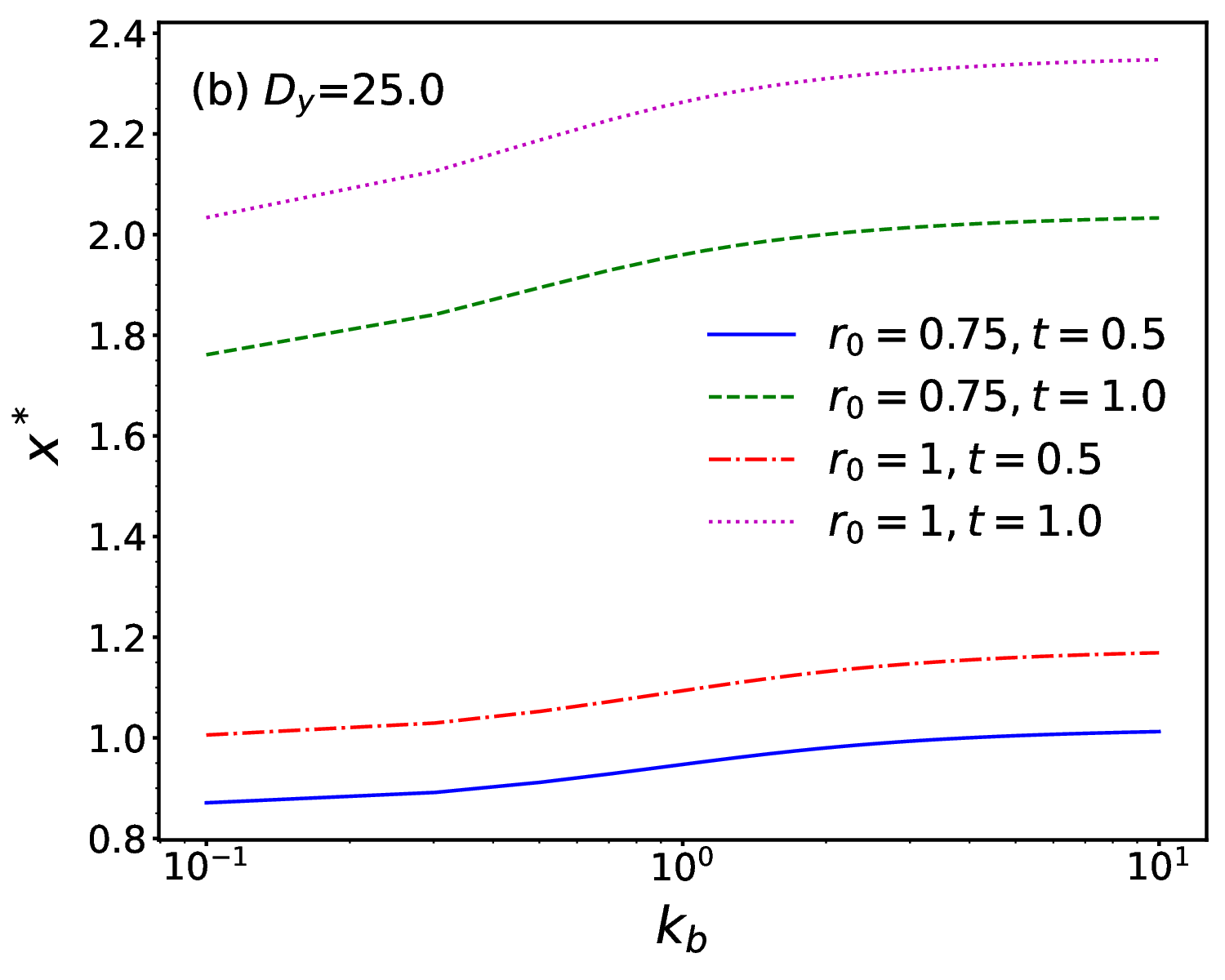}
    \end{subfigure}
\caption{Log-log plot of cross-over point $|x^*|$ [Eq. (\ref{xstar})] versus $k_b$ at two different times and for two values of resetting rate $r_0.$  Panels (a) and (b) correspond to the regimes $D_y<D_y^c$ and $D_y>D_y^c$, respectively, with $D_y^c = 11.0$.} \label{fig:xstar}
\end{figure*}

To understand how the distribution relaxes to steady state over all positions as time progresses, we employ the technique discussed in Ref. \cite{Evans2020}. We rescale the time as $\tau=\omega t$ and rewrite the PDF in Eq. (\ref{last_renewal_equation}) using Eq. (\ref{pdf_px4}) as
\begin{align}
P_r(x,t)= \frac{\exp\left(-t \Omega(1,t)\right)}{\sqrt{4\pi \xi_{\omega}(1,t)}}+ \frac{r_0 t}{\sqrt{4\pi}} \int_{0}^{1} d\omega\, \frac{\exp\left(-t \Omega(\omega,t)\right)}{\sqrt{4\pi \xi_{\omega}(\omega,t)}},\label{saddle_1}
\end{align}
where 
\begin{subequations}
\begin{align}
 &  \Omega(\omega,t)=r_0 \omega+\frac{x^2}{4t \xi_{\omega}(\omega,t)} \nonumber\\
 & \xi_{\omega}(\omega,t) = \nonumber\\
 &\alpha_1+\alpha_2 \omega t + \alpha_{11} e^{-k_b(N+\gamma) \omega t} +\alpha_{12} e^{-2k_b(N+\gamma) \omega t}.
\end{align}
\end{subequations}
 At long times, the PDF given by Eq. (\ref{saddle_1}) is predominantly determined by the second term on the RHS. Using the saddle-point method, the integration in Eq. (\ref{saddle_1}) can be performed by minimizing the large deviation function $ \Omega(\omega,t),$  which occurs at $\omega^*.$  Mathematically, this is given by  $\frac{\partial}{\partial\omega}\Omega(\omega^*,t)=0.$  From this, we can find a crossover point $x^*$ across which two regimes can be identified. Thus, one may obtain
 \begin{align}
&x^*=\nonumber\\
&\pm \sqrt{\frac{4 r_0 \xi_{\omega}(1,t)^2}{\alpha_2-k_b(N+\gamma)(\alpha_{11} e^{-k_b(N+\gamma) t} +2\alpha_{12} e^{-2k_b(N+\gamma) t})}}.\label{xstar}    
\end{align}
At a given time $t$, the density relaxes to steady state in the region  $|x|<x^*,$ while it remains time-dependent in the region $|x|>x^*.$ As time progresses, $x^*(t)$ moves further away from $x=0.$ Figure \ref{fig:xstar} shows that, at any given time $t$, $x^*(t)$
 occurs at a comparatively shorter distance for higher values of $k_b$ when $D_y<D_y^c$, and at a longer distance when $D_y>D_y^c.$
 In the small and  large $k_b$ limits, $x^*(t)$ can be simplified to $x^*(t) \sim \pm 2\sqrt{D_x r_0} t$ and $x^*(t) \sim \pm 2\sqrt{\alpha_2 r_0} t$, respectively, implying a linear growth of $x^*$ with time. 

\begin{figure}
\centering
\includegraphics[width=1\linewidth]{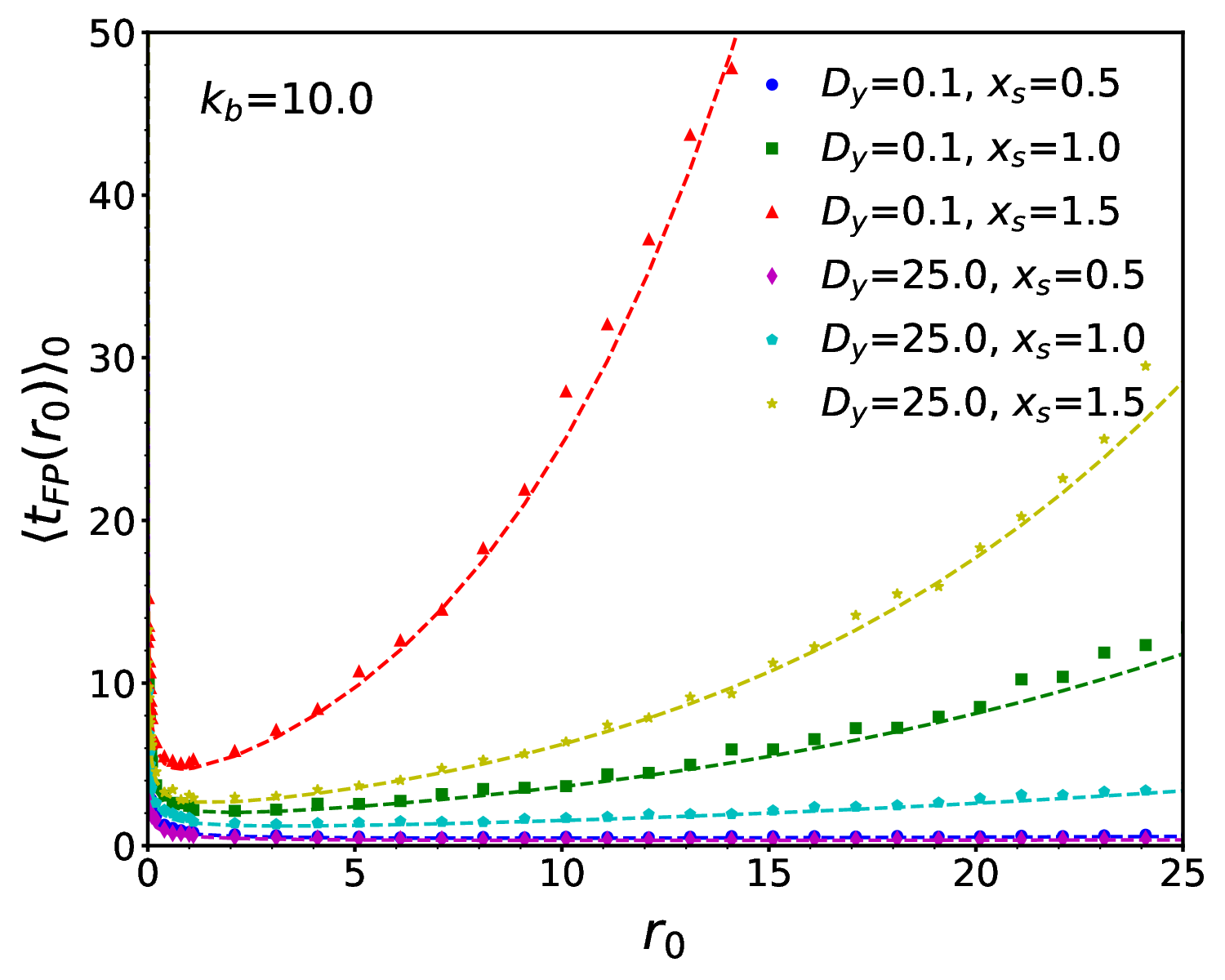}
\caption{Plot of the MFPT versus the resetting rate $r_0$ for three different target positions. The lines represent semi-analytical results for the MFPT derived from Eq. (\ref{tfp_free_0}) using Eq. (\ref{pdf_px4_0}), showing good agreement with the simulation data (symbols). Other parameters are the same as those in Fig. \ref{fig:msd}.} \label{fig:meantime_kb_xs1}
\end{figure}

 \begin{figure*}[htp]
    \centering
    \begin{subfigure}[b]{0.495\textwidth}
    \caption{}
        \centering
        \includegraphics[width=\textwidth]{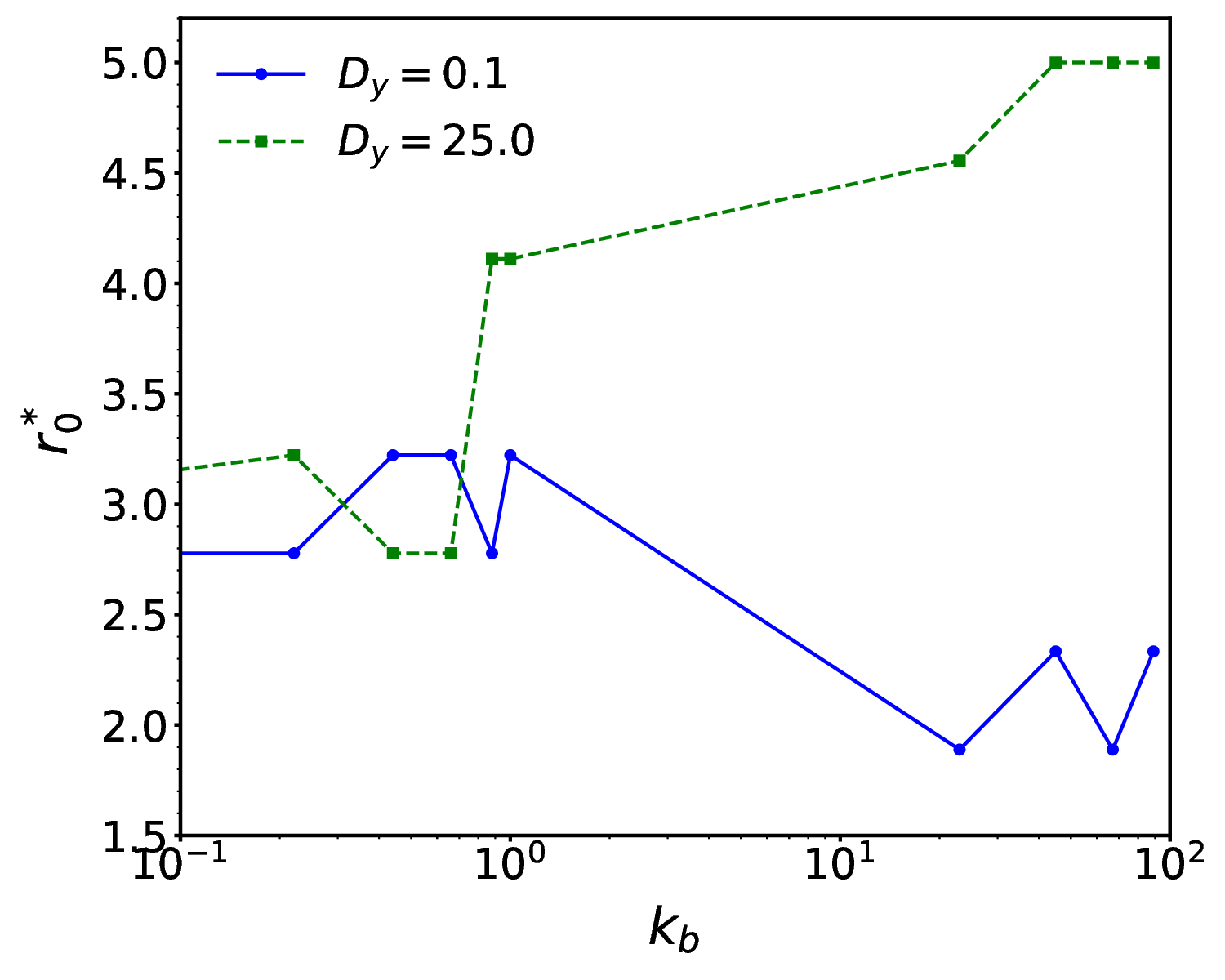}
    \end{subfigure}
    \hfill
    \begin{subfigure}[b]{0.495\textwidth}
    \caption{}
        \centering
        \includegraphics[width=\textwidth]{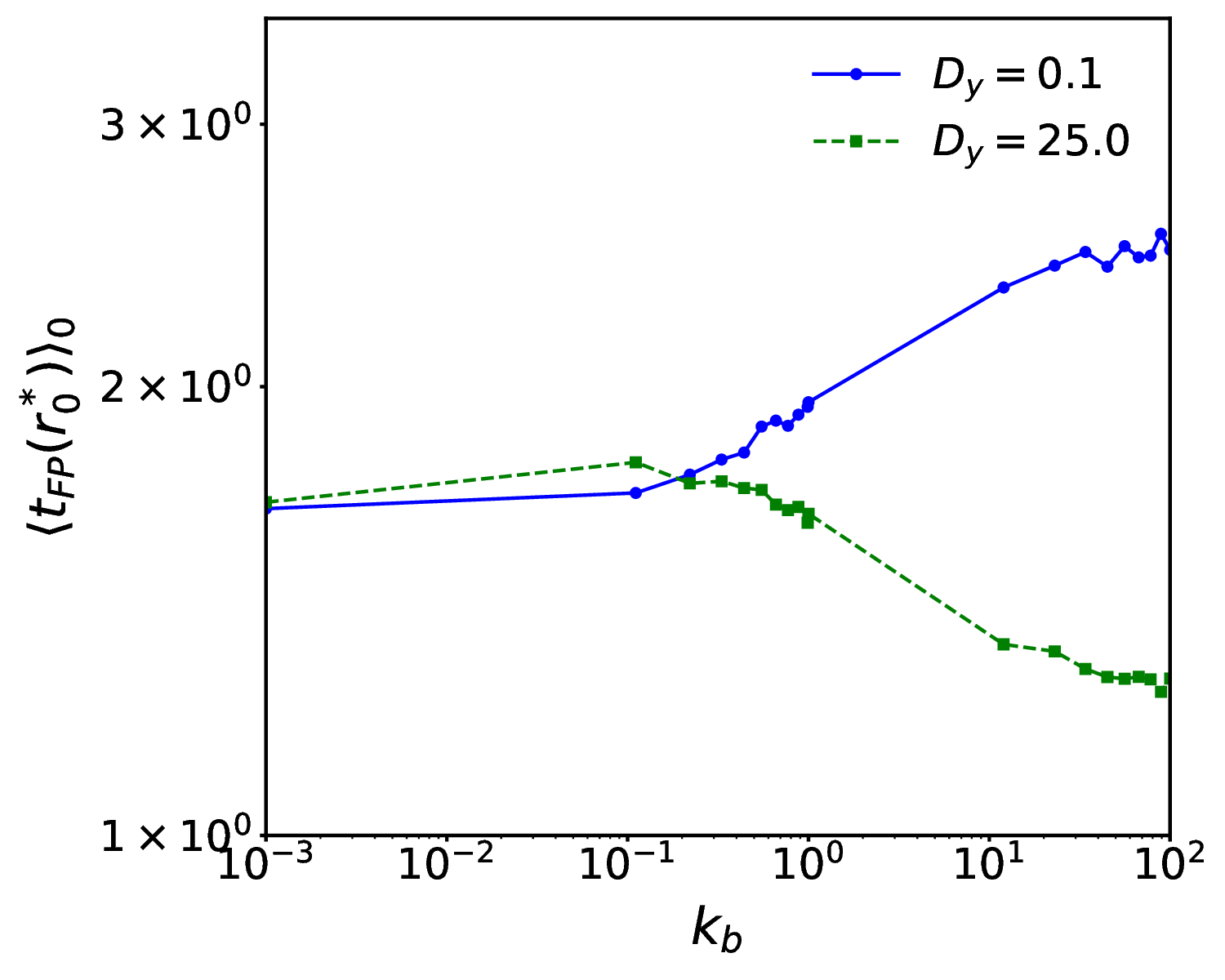}
    \end{subfigure}
\caption{(a)  Optimal resetting rate, $r_0^*$, extracted from the simulation data in Fig. \ref{fig:meantime_kb_xs},  plotted as a function of $\text{log}(k_b)$. (b) The MFPT at the optimal rate $r_0^*$ is shown as a function of  $k_b$ in a log-log scale. In both panels, the two curves represent the regimes $D_y<D_y^c$ and $D_y>D_y^c$, where $D_y^c=11.0.$} \label{fig:rstar_kb}
\end{figure*}

\section{First-passage times at different $k_b$ values \label{appen3}}

The first-passage density under resetting in the Laplace domain $s$ can be expressed as \cite{Reuveni2016,Reuveni2017}
\begin{align}
 \widetilde{f}_r(s)=\frac{\widetilde{f}_0(s+r_0)}{\frac{s}{s+r_0}+\frac{r_0}{s+r_0}\widetilde{f}_0(s+r_0)},    
\end{align}
where $\widetilde{f}_0(s)$ is the Laplace transform of the first-passage density in the absence of resetting. Thus, the mean first-passage time can be calculated as
\begin{align}
 \langle t_{FP}(r_0)\rangle=-\frac{\partial}{\partial s}\widetilde{f}_r(s)\Big|_{s=0}=\frac{1}{r_0}\frac{1}{\widetilde{f}_0(r_0)}-\frac{1}{r_0}.   
\end{align}
The first-passage density without resetting is known and is given by \cite{Janakiraman2017,goswami2021}
\begin{align}
\widetilde{f}_0(s)=\frac{ \widetilde{P}_0(x_s,s|x(0),0)}{ \widetilde{P}_0(x(0),s|x(0),0)},\label{tpf_noreset}
\end{align}
where $\widetilde{P}_0(x_s,s|x(0),0)$ is the Laplace transform of the reset-free propagator. Strictly speaking, the above relation is valid only for Markovian processes. In our case, the overall process is non-Markovian, which may lead to some discrepancies with the simulation data in certain parametric regimes [e.g., see Fig. \ref{fig:meantime_kb_xs}  for large values of $r_0$]. Nevertheless, it provides a complementary perspective to the simulation results, aligning well with the general trends observed.

In general, the MFPT of the tracer under resetting at a rate $r_0$ in the presence of a target located at position $x = x_s$, assuming its  initial position at  $x(0) = 0$, can be expressed as \cite{goswami2021}
\begin{align}
\langle t_{FP}(r_0)\rangle_0=\frac{1}{r_0}\Bigl(\frac{ \widetilde{P}_0(0,r_0|0,0)}{ \widetilde{P}_0(x_s,r_0|0,0)}-1\Bigr).\label{tfp_free}
\end{align}
Using the propagator of a free particle given by  Eq. (\ref{pdf_px4}), one can compute the Eq. (\ref{tfp_free}). 
However,  Eq. (\ref{tfp_free})  can be exactly determined only in the limits  $k_b \ll 1,$ and $k_b \gg 1,$ using the Laplace transform of  the reset-free propagators obtained from Eqs. (\ref{pdf_reset_st_approx_thermal})-(\ref{pdf_reset_st_approx_kbl}). In the limit $k_b \gg 1$, by virtue of Eq. (\ref{pdf_reset_st_approx_kbl}), the MFPT can be approximated to
\begin{align}
 T_1=\langle t_{FP}(r_0)\rangle_0 \approx   \frac{1}{r_0}\left[e^{x_s\sqrt{\frac{r_0}{\alpha_2}}}-1\right]. \label{tfp_free_approx} 
\end{align}
In the opposite limit  $k_b \ll 1$, we recover the typical result for a thermal bath, \begin{align}
 T_0=\langle t_{FP}(r_0)\rangle_0 \approx   \frac{1}{r_0}\left[e^{x_s\sqrt{\frac{r_0}{D_x}}}-1\right]. \label{tfp_free_approx_0} 
\end{align}
Here, we define a diffusive timescale $t_D$ as $t_D=x_s^2/D_x$ and assume that $\alpha_2 \sim \mathcal{O}(D_x)$. From Eqs. (\ref{tfp_free_approx})–(\ref{tfp_free_approx_0}), we can derive the following limiting cases for the ratio $T_1/T_0$:
\begin{align}
\frac{T_1}{T_0} 
& \approx  \begin{cases}
& \sqrt{\frac{D_x}{\alpha_2}},\; r_0 \ll t_D^{-1} \nonumber\\
&  e^{x_s\left[\sqrt{\frac{r_0}{\alpha_2}}-\sqrt{\frac{r_0}{D_x}}\right]},\; r_0 \gg t_D^{-1}.
\end{cases}\\ \label{T10}
\end{align}
In both limits, we find that  $T_1=T_0$ when $\alpha_2=D_x$. For $D_x>\alpha_2$, we have $T_1>T_0$, and for $D_x<\alpha_2$, one has $T_1<T_0$. 

\begin{figure*}[htp]
    \begin{subfigure}{0.495\textwidth} 
    \caption{}
    \centering
      \includegraphics[width=\textwidth]{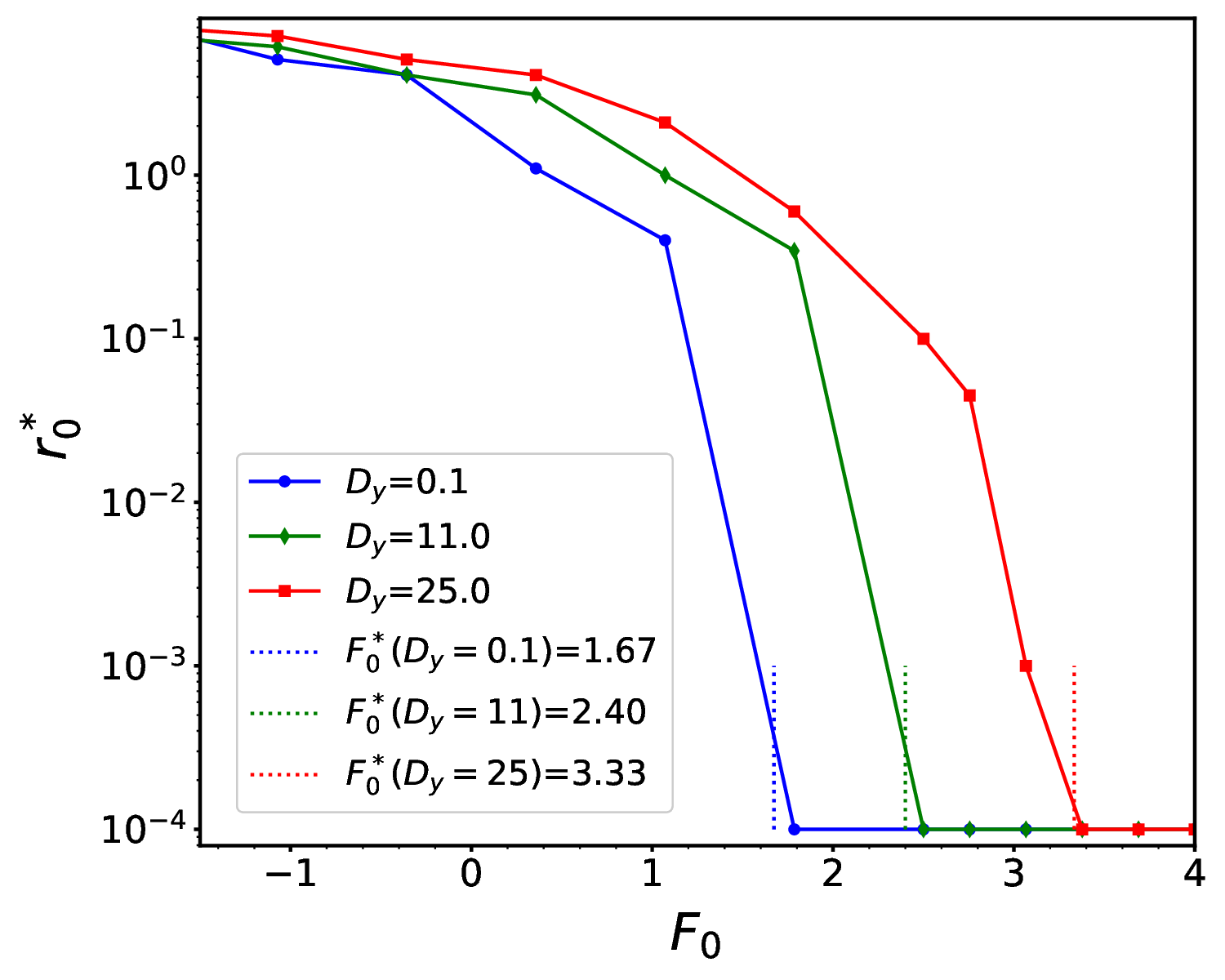}
    \end{subfigure}
     \begin{subfigure}{0.495\textwidth} 
     \caption{}
     \centering
     \includegraphics[width=\textwidth]{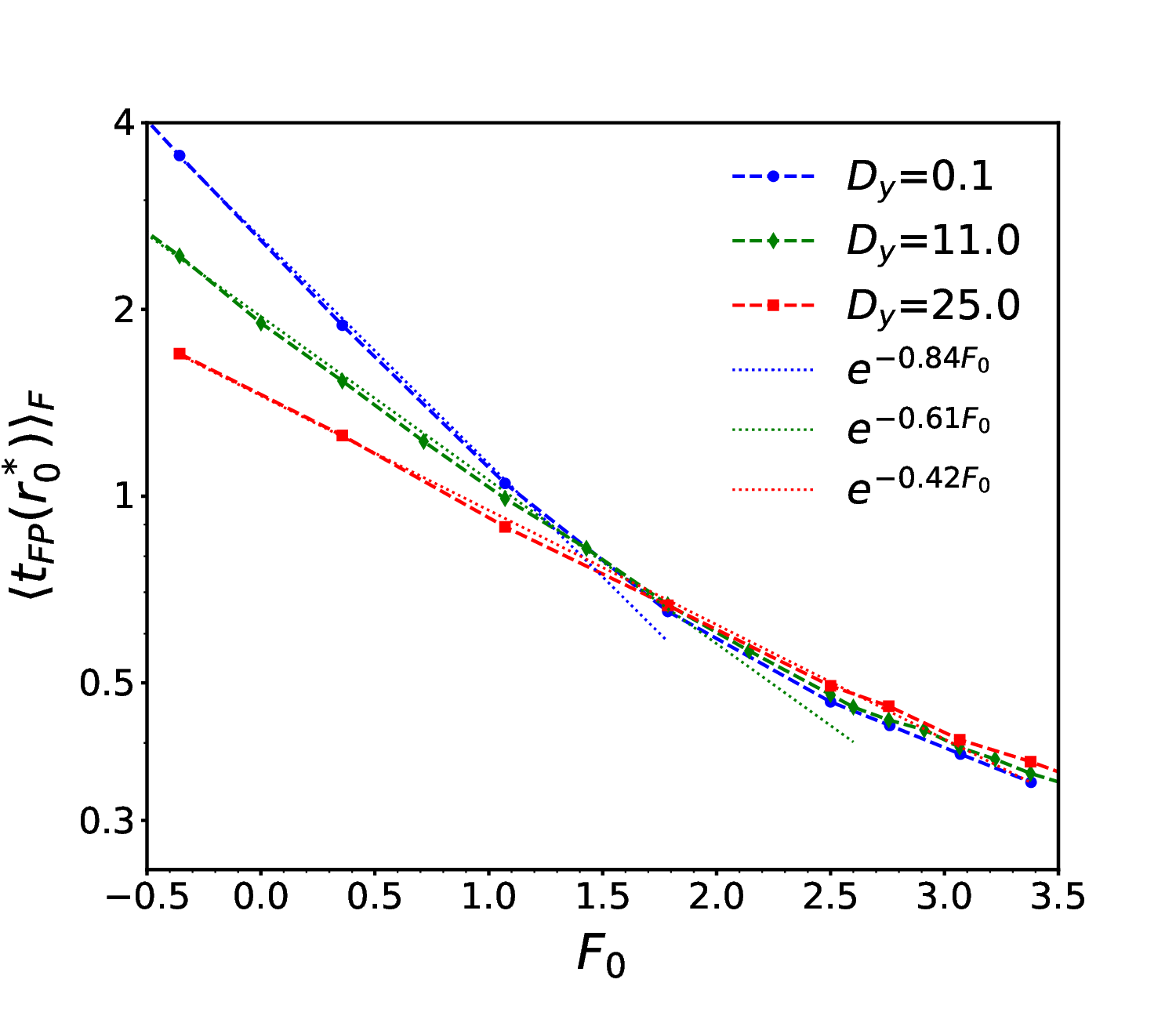}
    \end{subfigure}
\caption{(a) Logarithmic plot of the optimal resetting rate $r_0^*$ obtained from the simulation data plotted in Fig.  \ref{fig:meantime_f0}, as a function of the external force $F_0$  for three different values of  $D_y,$ representing the regimes $D_y < D_y^c$, $D_y = D_y^c$, and $D_y > D_y^c$, where $D_y^c = 11.0$. The analytical values of the critical force, $F_0^*$, obtained from Eq. (\ref{critical_force}), are indicated by dashed vertical lines. (b) Logarithmic plot of the MFPT at the optimal resetting rate $r_0^*$ as a function of $F_0$, showing  exponential decay fits.} 
\label{fig:rstar_f0}
\end{figure*}

The first-passage time for the tracer under the influence of an external force $F_0$ and in the presence of a target at position $x_s$ can be expressed as
\begin{align}
\langle t_{FP}(r_0)\rangle_F=\frac{1}{r_0}\Bigl(\frac{ \widetilde{P}_F(x(0),r_0|x(0),0)}{ \widetilde{P}_F(x_s,r_0|x(0),0)}-1\Bigr),\label{tfp_force}
\end{align}
where  $\widetilde{P}_F(x_s,r_0|x(0),0)$ is the Laplace transform of Eq. (\ref{pdf_px4_f0}), and  the  first-passage density without resetting in the presence of force $F_0$ is given by 
\begin{align}
  \widetilde{f}_F(s)= \frac{\widetilde{P}_F(x_s,r_0|x(0),0)}{ \widetilde{P}_F(x(0),r_0|x(0),0)}\label{tfp_force_fpt}
\end{align}
In the large $k_b$ limit, the Laplace transform of the propagator can be performed analytically.  So, by virtue of Eqs. (\ref{pf_noreset}) and (\ref{tfp_force_fpt}), we can write the density of FPT as 
\begin{align}
 \widetilde{f}_F(s)\approx e^{x_s\frac{\alpha_4}{2\alpha_2}-x_s\sqrt{\left(\frac{\alpha_4}{2\alpha_2}\right)^2+\frac{s}{\alpha_2}}}.   \label{tpf_noreset_approx}
\end{align}
Using the above results, one can approximate the MFPT in the presence of resetting and force $F_0,$ as 
\begin{align}
T_{F1}=\langle t_{FP}(r_0)\rangle_F \approx \frac{1}{r_0}\left[e^{-x_s\frac{\alpha_4}{2\alpha_2}+x_s\sqrt{\left(\frac{\alpha_4}{2\alpha_2}\right)^2+\frac{r_0}{\alpha_2}}}-1\right]. \label{tfp_force_approx}
\end{align}
This expression resembles the MFPT for a diffusing particle under a linear potential in a thermal bath, as discussed in \cite{Ahmad2019}, but here it includes rescaled diffusivity and force parameters.
For $k_b\ll 1,$ corresponding to the thermal bath case, the MFPT can be approximated using Eq. (\ref{pf_noreset_0}) as
\begin{align}
T_{F0}=\langle t_{FP}(r_0)\rangle_F \approx \frac{1}{r_0}\left[e^{-x_s\frac{F_0}{2D_x}+x_s\sqrt{\left(\frac{F_0}{2D_x}\right)^2+\frac{r_0}{D_x}}}-1\right], \label{tfp_force_approx_thermal}
\end{align}
which matches the result found in Ref. \cite{Ahmad2019}.

Let us define an additional timescale, the drift timescale, as $t_{F}=x_s/|F_0|.$ Applying Eqs. (\ref{tfp_force_approx})–(\ref{tfp_force_approx_thermal}), the following approximation is derived 
 \begin{align}
\frac{T_{F1}}{T_{F0}} \approx &
\begin{cases}
 &  \frac{\Theta_1}{\Theta_0},\; r_0 \ll t_D^{-1}\,\text{and }  r_0 \ll t_F^{-1}   \nonumber\\
 & e^{x_s(\Theta_1-\Theta_0)},\; r_0 \gg t_D^{-1}\,\text{and }  r_0 \gg t_F^{-1}
\end{cases},
 \end{align}
where $\Theta_1=-\frac{\alpha_4}{2\alpha_2}+\sqrt{\left(\frac{\alpha_4}{2\alpha_2}\right)^2+\frac{r_0}{\alpha_2}}$, and $\Theta_0=-\frac{F_0}{2D_x}+\sqrt{\left(\frac{F_0}{2D_x}\right)^2+\frac{r_0}{D_x}}.$ From the above, one gets $T_{F1}=T_{F0}$ when $\Theta_1=\Theta_0,$ which implies 
\begin{align}
 D_y=D_x(2\gamma+N)+\frac{F_0}{2r_0}(\gamma+N)\left[F_0+\sqrt{F_0^2+4 D_x r_0}\right].\label{Dy_regime2}  
\end{align}
In the presence of external force, Eq. (\ref{Dy_regime2}) defines  $D_y^c$, which characterizes regime II. As a consistency check, setting $F_0=0$ recovers the force-free result: $D_y^c=D_x(2\gamma+N).$ In the limit $F_0 \rightarrow -\infty,$ one can approximate $D_y^c \approx D_y^{nc}= D_x \gamma.$ For very large positive values of $F_0,$ $D_y^c$ scales as $D_y^c \propto F_0^2.$

\section{Useful Fourier and Laplace Transforms \label{appen4}}
The Fourier transform of the reset-free propagator in the presence of an external force $F_0$ in the large $k_b$ limit can be obtained from Eq. (\ref{pdf_px4_f0}), and it has the following form 
\begin{align}
\hat{P}(q,t) =e^{-(c_1+c_2 t) q^2+i\,(c_3+c_4 t) q},
\end{align}
where $P(x,t)=\frac{1}{2\pi}\int_{-\infty}^{+\infty}dq\,e^{-i q x}\,\hat{P}(q,t).$
The Laplace transform of the above can be expressed as
\begin{align}
 \Tilde{\hat{P}}(q,s)&  = \int_{0}^{\infty}dt\,e^{-s t} \,e^{-(c_1+c_2 t) q^2+i\,(c_3+c_4 t) q} \nonumber\\
&=\frac{e^{-c_1 q^2+i c_3 q}}{c_2 q^2-i c_4 q+s}.  \label{P(q,s)} 
\end{align}
\begin{widetext}
Upon the inverse Fourier transform of Eq. (\ref{P(q,s)}), one obtains the Laplace transform of the function $P(x,t)$ for $c_1>0$, and it reads 
\begin{align}
&\Tilde{P}(x,s)=\int_{-\infty}^{+\infty} dx'\,\Tilde{P}_1(x',s)\,\Tilde{P}_2(x-x',s)=\frac{1}{2 \sqrt{c_4^2 + 4 c_2 s}} \nonumber\\
&\times\Bigl[\exp\left(\frac{c_1 c_4 (c_4 - \sqrt{c_4^2 + 4 c_2 s}) + c_2 (2 c_1 s + (c_4-\sqrt{c_4^2 + 4 c_2 s}) (x-c_3))}{2 c_2^2}\right)
\operatorname{Erfc}\left(-\frac{c_1 (c_4 - \sqrt{c_4^2 + 4 c_2 s}) + c_2 (x-c_3 )}{2 \sqrt{c_1} c_2}\right) \nonumber\\
& +
\exp\left(\frac{c_1 c_4 (c_4 + \sqrt{c_4^2 + 4 c_2 s}) + c_2 (2 c_1 s + (c_4 + \sqrt{c_4^2 + 4 c_2 s}) (x-c_3))}{2 c_2^2}\right)
\operatorname{Erfc}\left(\frac{c_1 (c_4 + \sqrt{c_4^2 + 4 c_2 s}) + c_2 (x-c_3)}{2 \sqrt{c_1} c_2}\right)\Bigr],  \label{p_f} 
\end{align}
where 
\begin{equation}
\begin{array}{cc}
\begin{aligned}
\Tilde{P}_1(x,s) &=\frac{1}{2\pi}\int dq\, \frac{e^{-i q x}}{c_2 q^2-i c_4 q+s} \nonumber\\ 
 &=\frac{e^{-\frac{|x|}{2c_2}\sqrt{c_4^2+4c_2 s}+\frac{c_4\,x}{2c_2}}}{\sqrt{c_4^2+4 c_2 s}},
\end{aligned}
& \quad \quad
\begin{aligned}
 \Tilde{P}_2(x,s) &=\frac{1}{2\pi}\int dq\,e^{-i q x}\,e^{-c_1 q^2+i c_3 q} \nonumber\\
 & =\sqrt{\frac{1}{4\pi c_1}}\,e^{-\frac{(x-c_3)^2}{4c_1}}.
\end{aligned}
\end{array}
\end{equation}
\end{widetext}

\section{Simulation details \label{appen5}}
We solve Eqs. (\ref{langevin-tracer_reset1})-(\ref{langevin-bath_reset}) for the free particle case and Eqs. (\ref{langevin-tracer_f0_reset1})-(\ref{langevin-tracer_f0_reset2}) for the forced particle case using the Euler-Maruyama method with a time step of $\Delta t=5 \times 10^{-4}$ for a total simulation time of $t=100$. Simulations are carried out on independent $10^3$ trajectories to ensure statistical accuracy. To determine the mean first-passage time, we apply the aforementioned equations to both cases, with the condition that whenever the tracer reaches the target or sink at $x=x_s$, it is absorbed, thereby completing the process. The time taken for this event is recorded as the first-passage time. This procedure is repeated across $10^3$ independent trajectories, and the mean first-passage time is then computed for various resetting rates by averaging over these trajectories.

For all the cases, the initial position of the tracer $x$ is set to $x(0)=0,$ while the initial positions of the auxiliary variables $y$ are drawn from a normal distribution given by $P(y(0))$,  where  $P(y(0))=\sqrt{\frac{\gamma k_b}{2\pi D_y}} \exp\left(-\frac{\gamma k_b}{2D_y}y^2(0)\right).$ The specific parameters used in the simulations are  $N=1,\, D_x=1,\, \text{and}\, \gamma=5.$ For the case where both the tracer and bath particles are   in contact with the same thermal reservoir, the condition $D_y=D_y^{nc}=\gamma D_x$ is applied. To model a nonequilibrium bath, we consider several values of $D_y,$ as specified in the respective plots. Simulations are performed for various values of the resetting rate $r_0$ and the coupling constant $k_b,$ as indicated in the respective plots. All parameters are expressed in their respective units.




\begin{thebibliography}{99}
\bibitem{van1992stochastic} N.~G. Van~Kampen  1991 {\em Stochastic processes
in physics and chemistry \/} (Amsterdam: North-Holland)

\bibitem{frey2005brownian} E. Frey and K. Kroy 2005 {\em Ann. Phys., Lpz. \/} {\bf 14} 20

\bibitem{Zwanzig2002} R. Zwanzig 2001 {\em non-equilibrium Statistical
Mechanics \/}  (Oxford: Oxford University Press)

\bibitem{Kubo2012} R. Kubo, M. Toda, and N. Hashitsume 2012 {\em Statistical
Physics II: non-equilibrium Statistical Mechanics\/}  (Berlin: Springer)

\bibitem{Metzler2000} R. Metzler and J. Klafter 2000 {\em Phys. Rep. \/} {\bf 339} 1

\bibitem{Ferrer2021} B. R. Ferrer, J. R. Gomez-Solano,  and  A. V. Arzola 2021 {\em Phys. Rev. Lett.\/} {\bf 126} 108001

\bibitem{Hofling2013} F. H\"{o}fling and T. Franosch  2013 {\em Rep. Prog. Phys. \/} {\bf 76} 046602

\bibitem{Mitterwallner2020} B. G. Mitterwallner, C. Schreiber, J. O. Daldrop, J. O.
R\"{a}dler, and R. R. Netz 2020{\em Phys. Rev. E \/} {\bf 101} 032408 

\bibitem{Makarov2013} D. E. Makarov 2013 {\em J. Chem. Phys. \/} {\bf 138} 014102 

\bibitem{Kappler2018} J. Kappler, J. O. Daldrop, F. N. Brünig
, M. D. Boehle, and R. R. Netz 2018 {\em J. Chem. Phys. \/} {\bf 148} 014903

\bibitem{Mizuno2007} Mizuno D, Tardin C, Schmidt C~F, and  MacKintosh F~C
2007 {\em Science\/} {\bf 315} 370

\bibitem{Godec2014} A. Godec, M. Bauer, and R. Metzler 2014 {\em New J. Phys.\/}
{\bf 16} 092002

\bibitem{Joo2022} Y. Kim, S. Joo, W.~K. Kim, and J.-H. Jeon  2022 {\em
Macromolecules\/} {\bf 55} 7136

\bibitem{Theeyancheri2022} L. Theeyancheri, R. Sahoo, P. Kumar, and R. Chakrabarti  2022 {\em ACS Omega\/} {\bf 7} 33637

\bibitem{Faris2009} M~D~El Alaoui Faris , D. Lacoste, J. P\'ecr\'eaux,  J.~F. Joanny, J. Prost, and P. Bassereau 2009 {\em  Phys. Rev. Lett.\/} {\bf 102} 038102

\bibitem{Park2010} Y. Park,  C.~A. Best, T. Auth,  N.~S. Gov, S.~A. Safran, G. Popescu, S.  Suresh, and M.~S. Feld  2010 {\em Proc. Natl.  Acad. Sci. USA\/} {\bf 107}
(4) 1289

\bibitem{Crisanti2012} A. Crisanti, and A. Puglisi, and D. Villamaina 2012 {\em Phys. Rev. E\/} {\bf 85} 8802

\bibitem{Fodor2014} {\'E}. Fodor ,K. Kanazawa, H. Hayakawa, P. Visco, and F. Van
Wijland 2014 {\em Phys. Rev. E\/} {\bf 90} 061127

\bibitem{Hiraiwa2018} T Hiraiwa and R~R Netz  2018 {\em Europhys. Lett. \/}
{\bf 123} 58002

\bibitem{Netz2018}  R. Netz 2018 {\em J. Chem. Phys.\/} {\bf 148} 185101

\bibitem{Goswami2023} K. Goswami and R. Metzler 2023 {\em Soft Matter\/} {\bf 19} 8802

\bibitem{Wang2021} M. Wang M, K. Zinga, A. Zidovska, and  A.~Y. Grosberg 2021 {\em
Soft Matter \/} {\bf 17} 9528

\bibitem{Guevara-Valadez2023}  C. A. Guevara-Valadez, R. Marathe, and J. R. Gomez-Solano 
2023 {\em Physica A\/} {\bf 609} 128342

\bibitem{Caprini2024} L. Caprini, A. Ldov, R. K. Gupta, H. Ellenberg,
R. Wittmann, H. L\"{o}wen, and C. Scholz 2024 {\em Commun. Phys. \/} {\bf 7} 52

 \bibitem{Ginot2022} Ginot F, Caspers J, Krüger M, and Bechinger C 2022 {\em Phys. Rev. Lett.\/} {\bf 128} 028001

 \bibitem{Das2023} B. Das, S. Paul, S. K. Manikandan, and A. Banerjee 2023 {\em New J. Phys.} {\bf 25} 093051
 
 \bibitem{Filliger2007} R. Filliger and P. Reimann 2007 {\em Phys. Rev. Lett.\/} {\bf 99} 230602

\bibitem{Berut2016} A. Bérut, A. Imparato, A. Petrosyan, and S. Ciliberto 2016 {\em Phys. Rev. Lett. \/}{\bf 116} 068301

\bibitem{Medina2018} E. Medina, R. Satija, and D. E. Makarov,2018 {\em J. Phys. Chem. B \/} {\bf 122} 11400

\bibitem{Hippel1989} P.H. von Hippel, O.G. Berg 1989 {\em J. Biol. Chem. \/} {\bf 264}  675

\bibitem{Bauer2012} M. Bauer and R. Metzler 2012 {\em Biophys. J. \/} {\bf 102}  2321

\bibitem{Li2009} F. Li, JZ. Tsien 2009 {\em N Engl J Med. \/} {\bf 361} 302

\bibitem{Northrup1982}S. H. Northrup, F. Zarrin, and J. A. McCammon {\em J. Phys. Chem.\/} {\bf 86} 2314

\bibitem{Fauchald2003} P. Fauchald and T. Tveraa 2003 {\em Ecology \/} {\bf 84} 282

\bibitem{Monasterio2011} C. Mejia-Monasterio, G. Oshanin, and G. Schehr 2011 {\em J. Stat. Mech. \/}  P06022 

\bibitem{Noton19711} D. Noton and L. Stark 1971 {\em Science \/} {\bf 171} 308
 
\bibitem{Noton19712} D. Noton and L. Stark 1971 Vis. Res. 1971; 11(9):929–IN8.

\bibitem{Henderson2003} J. M. Henderson 2003 {\em  Trends Cogn. Sci. \/} {\bf 7}  498

\bibitem{Redner2001} S. Redner 2001 {\em A Guide to First-Passage Processes \/}  (Cambridge University Press, Cambridge)

\bibitem{Evans2020}Evans, M R and Majumdar, S N, and Schehr, G 2020 {\em J. Phys. A: Math. Theor.\/}
{\bf 53} 193001

\bibitem{Evans2011} M. R. Evans and S. N. Majumdar 2011 {\em Phys. Rev. Lett.\/}  {\bf 106} 160601

\bibitem{Friedman2020} O. Tal-Friedman, A. Pal, A. Sekhon, S. Reuveni, and Y.
Roichman 2020 {\em J. Phys. Chem. Lett. \/}  {\bf 11} 7350

\bibitem{Evans2011j} M. R. Evans and S. N. Majumdar 2011 {\em J. Phys. A \/} {\bf 44} 435001

\bibitem{Gupta2020} D. Gupta, C. A. Plata, A. Kundu, and A. Pal 2020 {\em J. Phys. A: Math. Theor. \/} {\bf 54} 025003

\bibitem{Méndez2021} V. Méndez, A. Masó-Puigdellosas, T. Sandev, and D. Campos 2021
 {\em Phys. Rev. E \/} {\bf 103} 022103

\bibitem{Kusmierz2014} L. Kusmierz, S. N. Majumdar, S. Sabhapandit, and G.
Schehr 2014 {\em Phys. Rev. Lett. \/} {\bf 113} 220602

\bibitem{Majumdar2018} S. N. Majumdar and G. Oshanin 2018 {\em J. Phys. A: Math. Theor.} {\bf 51} 435001

\bibitem{Vinod2022} D. Vinod, A. G. Cherstvy, W. Wang, R. Metzler, and I. M.
Sokolov 2022 {\em Phys. Rev. E} {\bf 105} L012106 

\bibitem{Evans2018} M. R. Evans and S. N. Majumdar 2018 {\em J. Phys. A: Math. Theor. \/} {\bf 51} 475003 

\bibitem{goswami2021} K. Goswami and R. Chakrabarti 2021 {\em  Phys. Rev. E \/}
{\bf 104} 034113

\bibitem{abdoli2021} I. Abdoli and A. Sharma 2021 {\em  Soft Matter \/}
{\bf 17} 1307

\bibitem{Gupta2014} S. Gupta, S. N. Majumdar, and G. Schehr 2014 {\em Phys. Rev. Lett. \/} {\bf 112} 220601 

\bibitem{Singha2023} T. Singha 2023 {\em Phys. Rev. E \/} {\bf 107} 044117

\bibitem{Bodrova2019} A. S. Bodrova, A. V. Chechkin, and I. M. Sokolov 2019 {\em Phys. Rev. E} {\bf 100} 012119 

\bibitem{Bodrova20191} A. S. Bodrova, A. V. Chechkin, and I. M. Sokolov 2019 {\em Phys. Rev. E} {\bf 100} 012120 

\bibitem{Blossey2019} R. Blossey, H. Schiessel 2019 {\em J. Phys. A} {\bf 52}  085601.

\bibitem{Reuveni2016} S. Reuveni  2016 {\em  Phys. Rev. Lett. \/}
{\bf 116} 170601

\bibitem{Ahmad2019} S. Ahmad, I. Nayak, A. Bansal, A. Nandi, and D. Das 2019 {\em  Phys. Rev. E \/}
{\bf 99} 022130

\bibitem{Biswas2023} A. Biswas, A. Pal, D. Mondal, and S. Ray 2023 {\em  J. Chem. Phys. \/} {\bf 159} 054111

\bibitem{Kimura2023} M. Kimura, T. Akimoto 2023 {\em J. Chem. Phys.} {\bf 159} 055102 

\bibitem{harbola2024} 
 U. Harbola 2024 {\em Phys. Rev. E\/} {\bf 109}, 064148.

\bibitem{Goychuk2009} I. Goychuk 2009 {\em Phys. Rev. E} {\bf 80} 046125 

\bibitem{goswami2019heat} K. Goswami  2019 {\em Phys. Rev. E\/} {\bf 99} 012112

\bibitem{Abbasi2023} A. Abbasi, R. R. Netz, and A. Naji 2023 {\em Phys. Rev. Lett.} {\bf 131} 228202
(2023).

\bibitem{Lahiri2024} S. Lahiri and  S. Gupta  2024 {\em  Phys. Rev. E \/}
{\bf 109} 014129

\bibitem{Kumari2024} A. Kumari, M. Samsuzzaman, A. Saha, and S. Lahiri 2024 {\em  Physica A \/}
{\bf 636} 129575

\bibitem{goswami2023j} K. Goswami and R. Metzler 2023 {\em  J. Phys. Complex. \/}
{\bf 4} 025005

\bibitem{goswami2019diffusion} K. Goswami and K.~L. Sebastian 2019 {\em
J. Stat. Mech.: Theory Exp.: Theory Exp.\/}  083501

\bibitem{Reuveni2017} A. Pal and S. Reuveni 2017 {\em  Phys. Rev. Lett. \/}
{\bf 118} 030603

\bibitem{Janakiraman2017} D. Janakiraman 2017 {\em  Phys. Rev. E \/}
{\bf 95} 012154

\end{thebibliography}
\end{document}